\documentclass[english,preprintnumbers,amsmath,amssymb,twocolumn]{revtex4}
\usepackage[T1]{fontenc}
\usepackage[latin9]{inputenc}
\usepackage{color}
\usepackage{float}
\usepackage{amsbsy}
\usepackage{amstext}
\usepackage{graphicx}
\usepackage{esint}
\usepackage{epstopdf}
\usepackage{hyperref}

\usepackage{amsmath}

\usepackage[normalem]{ulem}

\makeatletter

\@ifundefined{textcolor}{}
{%
 \definecolor{BLACK}{gray}{0}
 \definecolor{WHITE}{gray}{1}
 \definecolor{RED}{rgb}{1,0,0}
 \definecolor{GREEN}{rgb}{0,1,0}
 \definecolor{BLUE}{rgb}{0,0,1}
 \definecolor{CYAN}{cmyk}{1,0,0,0}
 \definecolor{MAGENTA}{cmyk}{0,1,0,0}
 \definecolor{YELLOW}{cmyk}{0,0,1,0}
 }

\usepackage{dcolumn}
\usepackage{bm}
\usepackage{array}

\def\ket{\rangle}
\def\bra{\langle}

\def\psiBpm{\psi_{\textrm{BS},\pm}}
\def\psiBp{\psi_{\textrm{BS},+}}
\def\psiBm{\psi_{\textrm{BS},-}}

\def\gstr{g_\textrm{str}}
\def\gw{g_\textrm{weak}}

\DeclareMathOperator{\sign}{sgn}

\DeclareMathOperator{\Abr}{A_\textrm{br}}

\newcommand{\re}{\textrm{Re}}
\newcommand{\im}{\textrm{Im}}

\newcommand{\Hbath}{H_\textrm{semi}}
\newcommand{\Hssh}{H_{\textrm{SSH}}}
\newcommand{\Hinf}{H_{{\textrm{SSH},\infty}}}

\newcommand{\beq}{\begin{equation}}
\newcommand{\eeq}{\end{equation}}
\newcommand{\beqa}{\begin{eqnarray}}
\newcommand{\eeqa}{\end{eqnarray}}

\makeatother

\usepackage{babel}

\begin{document}

\title{Weak-coupling bound states in semi-infinite topological waveguide QED}
\author{Savannah Garmon}
%
\affiliation{Department of Physics, 
Osaka Metropolitan University, 
3-3-138 Sugimoto, Osaka 558-8585, Japan}
\affiliation{Institute of Industrial Science, University of Tokyo, Kashiwa 277-8574, Japan}
\author{Gonzalo Ordonez}
\affiliation{Department of Physics and Astronomy, Butler University, Gallahue Hall, 4600 Sunset Avenue, Indianapolis, Indiana 46208, USA}
\author{Kenichi Noba}
\affiliation{Department of Physics, 
Osaka Metropolitan University, 
3-3-138 Sugimoto, Osaka 558-8585, Japan}

\begin{abstract}
A striking feature of cavity quantum electrodynamics is the existence of atom-photon bound states, which typically form when the coupling between the atom and its environment are strong enough that the atom can ``grab'' an emitted photon and re-absorb it, resulting in a virtual cloud surrounding the atom.  Here we will demonstrate the existence of bound states that instead form in the case of weak coupling.  Specifically, we show that when a quantum emitter is weakly coupled to a structured reservoir exhibiting topologically-protected surface states, hybridizations between these states and the emitter can form, resulting in mid-gap bound states.  We illustrate this using a semi-infinite extension of the Su-Schrieffer-Heeger (SSH) model as our reservoir. First, we diagonalize the bare semi-infinite SSH chain and reveal a winding number that predicts that only the edge state on the finite side of the chain survives the semi-infinite extension. Then, after coupling the quantum emitter to this end of the chain, we analyze the modified emitter spectrum and reveal the existence of bound states in three parameter regions. Two of these represent the usual strong-coupling bound states, while the third gives the weak-coupling bound states with eigenvalue appearing in the SSH band gap and which exhibit partial sublattice localization.  We demonstrate that oscillations between the weak-coupling bound states can be used to transfer the particle from the emitter into the lattice in a predictable and reversible manner.
\end{abstract}

\maketitle

\section{introduction}
\label{sec:intro}

The radiative properties of atoms (or more generally, quantum emitters) can be dramatically modified when placed in a structured photonic reservoir \cite{LNNB00}, such as a waveguide or a photonic crystal.
In the latter case, the periodic properties of the artificial crystal can induce a photonic band gap in which the propagation of light is suppressed over a range of frequencies \cite{Nature97,Yablonovitch87,John87}.  An atom or a defect impurity introduced in the crystal can then induce intriguing phenomena including atom-photon bound states \cite{Bykov75,Kurizki90,Lombardo14,CCCR16,SWGTC16,Zueco17} various schemes to induce superradiance and subradiance \cite{LonghiEPJ07,TudelaPRA17,Branczyk20} as well as non-trivial level splittings near a photonic band edge or threshold \cite{JohnPRL90,JohnPRB91,KKS94,John94,GNHP09,GarmonPRR21}.

The atom-photon bound state is particularly striking in that it results in a system that simultaneously exhibits stability and instability.  Essentially, the system can partly decay but asymptotically in time the atom-photon bound state remains occupied, resulting in ``fractional decay'' \cite{Bykov75,Lombardo14,KKS94,John94,GarmonPRR21,Schulman95}.  This occurs due to a form of photon localization: essentially, the atom can emit a photon, but the photon associated with the bound state is continuously reabsorbed, resulting in a virtual cloud surrounding the atom.  
Usually the atom-photon bound state occurs in the case of strong atom-field coupling, in which the atom can ``grab'' the escaping photon and pull it back.   Several examples of this type of bound state can be found in the literature \cite{Lei2011,BCP10,Liu2013,Longhi13,Tudela2014,YangAn2017,Thanopulos17}, including the experiment in Ref. \cite{NPoM2016} in which a single molecule embedded within a reservoir constructed from a nanoparticle-on-mirror geometry yields a split peak resonance associated with emitter-plasmon hybridized states \footnote{To give a more complete picture, we also mention the occurrence of bound states in certain systems that play a dominant role in the strong-coupling dynamics and yet still exist (although becoming less prominent) as one decreases the coupling into the weak-coupling regime \cite{Lombardo14,CCCR16,Zueco17,Thanopulos22}. (Almost equivalently, one can instead consider weak-coupling from the start, and then sweep the energy of the emitter from outside into the continuum \cite{GNHP09,Schulman95}.)  We call this the {\it persistent bound state} \cite{GNHP09}, which has been observed in the low-power transmission spectrum of a superconducting transmon qubit coupled to a photonic crystal in Ref. \cite{LiuHouck16} as well as an optical lattice experiment based on cold atoms in Ref. \cite{SKLS20}.  However, this state has some drawback in that its influence on physical quantities like excitation spectra \cite{GNHP09} and emitter survival probability \cite{Thanopulos22} would probably be rather humble in most weak-coupling scenarios.}.
However, we will provide an example of a counter-intuitive weak-coupling atom-photon bound state in what follows, which can be induced when the reservoir adopts properties analogous to those of a topological insulator.

Following the discovery of the optical quantum Hall effect \cite{Haldane2008,CAB08,WCJS08,WCJS09,Skirlo15,Hafezi13}, many researchers have focused on the role of topology in quantum-optical and photonic systems \cite{ModPhys19}.  Topologically-protected surface states have been observed, for example, in 1-D optically-induced photonic superlattices \cite{Malkova09} and 2-D helical waveguide arrays \cite{Szameit13}.  More recently, some researchers have begun to consider the influence of such systems when they act as a bath (a {\it topological structured reservoir}) for attached quantum emitters \cite{PRX1} that may interact with or modify the surface states, when present.
In this picture, the quantum emitter might be viewed as a means to probe or access the topological properties of the reservoir, or alternatively we could view this setup as a mechanism by which the emitter itself promulgates those properties.
 Previous studies along these lines
 have revealed interesting phenomena including one-sided bound states \cite{PRX1,BPCG19}, quasi-quantized decay rates \cite{Tudela2D23}, non-trivial symmetry-breaking for a non-Hermitian central potential \cite{GN21},
 and bound states that only appear in the weak-coupling regime, 
 which have been investigated numerically in  \cite{Zhang22} by Zhang, et al.


In this paper, we give a detailed analytical treatment of two different kinds of bound states that 
influence the dynamics most strongly in the weak-coupling case.  In the first scenario, we show that the emitter can essentially recreate the surface state in some circumstances.  In this case, the bound state also exists in the strong-coupling regime, but plays a greater role in the dynamics as the coupling is weakened.  
We then turn to a second scenario, which is the main subject of this paper, in which the emitter hybridizes with the surface state of the reservoir, giving rise to qualitatively distinct bound states that exist only in the weak-coupling regime.  These latter are qualitatively similar to the bound states from the Zhang, et al paper \cite{Zhang22} and further exhibit interesting dynamical features that are absent in the first scenario; however, we here provide a detailed analytical picture including a winding number that predicts the existence of these states. 

\begin{figure}
 \includegraphics[width=0.45\textwidth]{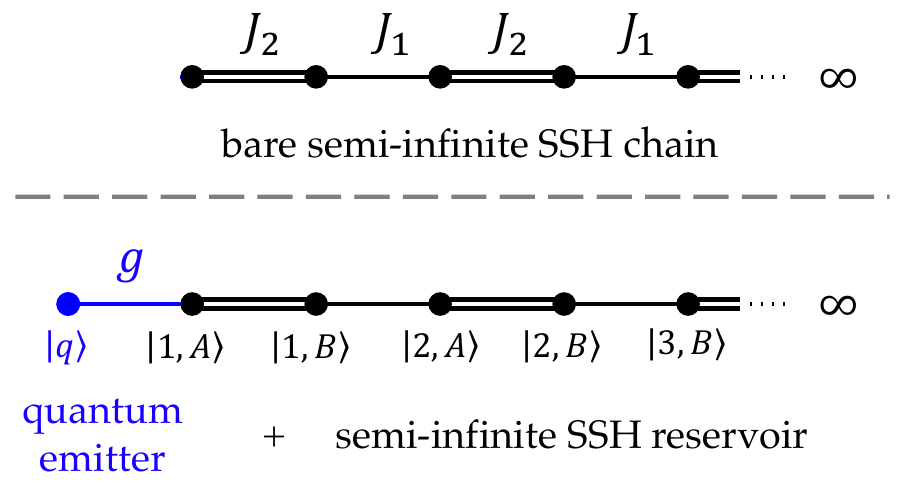}
\hfill
\\
\caption{ System geometry for (upper half) Eq. (\ref{semi.ham.2q}) Hamiltonian and (lower half) Eq. (\ref{1A.ham.2q}) Hamiltonian.
}
 \label{fig:geo}
 \end{figure}

We illustrate these concepts with a simple system in which a quantum emitter interacts with a 1-D topological bath. Our Hamiltonian is written as
\beq
  H =  g \left( \sigma_+ c_{1,A} + c_{1,A}^\dagger \sigma_- \right)  + \Hbath
\label{1A.ham.2q}
\eeq
in which $\sigma_+ = \sigma_-^\dagger = | q \ket \bra g |$ is the operator that transitions the quantum emitter from its internal ground state $| g \ket$ to its excited state $| q \ket$, via interaction with the bath,
while the bath Hamiltonian itself is given by a semi-infinite extension of the Su-Schrieffer-Heeger (SSH) Hamiltonian \cite{SSH1979,Asboth2016}, written as
\beqa
  \Hbath
  	& = & \sum_{n=1}^\infty \left[ J_2 \left( c_{n,A}^\dagger c_{n,B} + c_{n,B}^\dagger c_{n,A} \right) \right.
  				\nonumber	\\
  	& &	\left. + J_1 \left( c_{n+1,A}^\dagger c_{n,B} + c_{n,B}^\dagger c_{n+1,A} \right)	 \right]
	.
\label{semi.ham.2q}
\eeqa
Here $c_{n,A}^\dagger$ ($c_{n,B}^\dagger$) denotes the particle creation operator on the A (B) element in the $n$th unit cell, while $J_2$ denotes the intracell coupling and $J_1$ is the intercell coupling, as depicted in Fig. \ref{fig:geo}.
Notice from Eq. (\ref{1A.ham.2q}) we have assumed the emitter frequency is resonant with the cavity modes $c_{n, X}^\dagger$.  Hence, the remaining adjustable parameters in our model are the emitter-bath coupling $g$ and the two SSH bath parameters $J_1$ and $J_2$.  Further, we have employed the rotating wave approximation. 
This implies the particle number is conserved for the Hamiltonian in Eq. (\ref{1A.ham.2q}), and hence we will work in the single-particle picture from this point forward.  

In the next step, we work out the exact diagonalization procedure for the bare semi-infinite SSH chain in Sec. \ref{sec:diag}, which, as far as we are aware, has not previously appeared in the literature.  
We show that one of the two edge states of the finite SSH model survives the semi-infinite transition, resulting in an exponentially localized state with perfect sub-lattice localization in the topological configuration.  We further use the form of the continuum eigenstates to introduce a winding number associated with the surviving edge state.  
We extend this winding number analysis to the full Hamiltonian incorporating the quantum emitter in Sec. \ref{sec:1A} in order to predict the existence of bound states in various parameter regimes of that system, including the weak-coupling bound states.  
In Sec. \ref{sec:1A.spectrum} we introduce the Green's function in order to examine the generalized spectrum of the system, including anti-bound states, resonances and anti-resonances.
Applying the Green's function, we then briefly examine the survival probability dynamics of the initially-prepared state of the quantum emitter in Sec. \ref{sec:1A.dynamics}.  In particular, we demonstrate that the hybridized weak-coupling bound states can be used to direct the particle from the emitter into the bulk of the system in a predictable and reversible manner.  We provide a summary and closing comments in Sec. \ref{sec:conclusion}.




\section{Diagonalization of the bare semi-infinite SSH reservoir}\label{sec:diag}

\begin{figure}
 \includegraphics[width=0.45\textwidth]{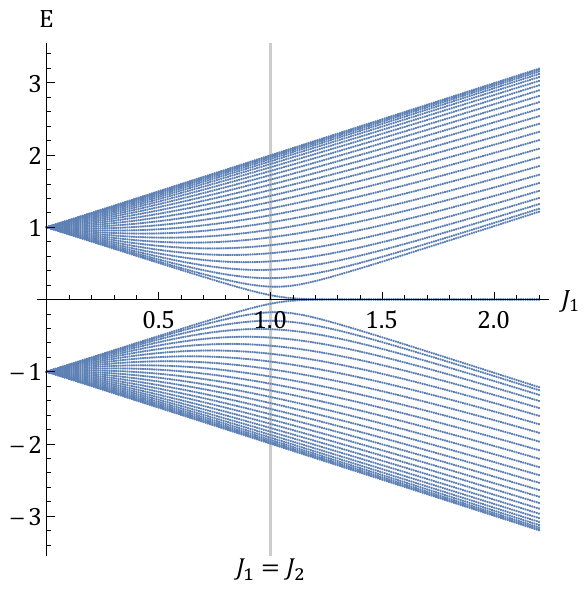}
\hfill
\\
\caption{ Energy eigenvalues of the finite SSH chain for $N = 26$ cells as a function of the parameter $J_1$, while $J_2 = 1$.  In the case $J_1 > J_2$ two eigenvalues split off from the bands and take on $E \approx 0$.  These are the two hybridized edge state eigenvalues.
}
 \label{fig:spec.ssh}
 \end{figure}

To guide our discussion below, we first briefly restate the key features of the original, finite SSH model \cite{SSH1979,Asboth2016}, with the Hamiltonian in the single-particle picture given by
\beqa
  \Hssh 
  &	= & J_2 \sum_{n = 1}^{N} \left( | n, A \ket \bra n, B | + | n, B \ket \bra n, A | \right)
				\\
  & & 	+ J_1 \sum_{n = 1}^{N-1} \left( | n, B \ket \bra n+1, A | + | n+1, A \ket \bra n, B | \right)
  				\nonumber	
\label{Hssh.defn}
\eeqa
in which 
$| n, A \ket = c_{n,A}^\dagger | \textrm{vac} \ket $ 
(or $| n, B \ket = c_{n,B}^\dagger | \textrm{vac} \ket $) denotes the state with the particle occupying the $n, A$ 
(or $n, B$) position on the lattice, and $| \textrm{vac} \ket $ denotes the vacuum.
This Hamiltonian can be shown to obey chiral sublattice symmetry in the form $\Sigma_C \Hssh \Sigma_C = - \Hssh$ in which $\Sigma_C = \hat{P}_A - \hat{P}_B$ is written in terms of the sublattice projection operators 
$\hat{P}_A = \sum_{n=1}^N | n, A \ket \bra n, A | $ and $\hat{P}_B = \sum_{n=1}^N | n, B \ket \bra n, B | $.  As a result, it is easy to show that every eigenstate $| E \ket$ with eigenvalue $E$ has a chiral partner given by $\Sigma_C | E \ket$ with eigenvalue $-E$.  This can be intuitively understood from the structure of the eigenstates themselves.  In the case that the intracell coupling $J_2$ is larger than the intercell coupling $J_1$ the system eigenstates can be written as linear combinations of dimer states formed by each $A, B$ pair within a given cell $n$.  In this case, the eigenvalues form two bands, as can be seen in Fig. \ref{fig:spec.ssh}.  This is the trivial case of the finite SSH model.

However, in the case $J_1 > J_2$, the relevant dimer states are instead comprised of the $A$ site in the $n$th cell and the $B$ site from the $n+1$ cell, such that these dimer states are now confined to the interior (``bulk'') of the chain, leaving one undimerized site on either end.  Two edge states hence form that are exponentially localized with strongest amplitude on these two unpaired sites, and which further exhibit sub-lattice localization.   However, note that the edge states themselves are not actually eigenstates in the finite model, as the associated eigenstates are instead formed from two linear combinations of these states.  These two eigenstates have energy $E \approx 0$ as can be seen in Fig. \ref{fig:spec.ssh} for $J_1 > J_2$.

In the following, we consider two different extensions of the SSH model.  First, we find the eigenstates for the case in which the chain extends to infinity in {\em both} directions.  We then obtain the continuum eigenstates for the semi-infinite chain as linear combinations from the infinite model.

\subsection{Infinite SSH chain: continuum eigenstates and diagonalized Hamiltonian}\label{sec:diag.inf}

The Hamiltonian for the the infinite SSH chain is written in the site representation as
\beqa
  \Hinf
	& = & \sum_{n=-\infty}^\infty \left[ J_2 \left( | n, A \ket \bra n, B | + | n, B \ket \bra n, A | \right) \right.
\label{inf.ham}   \\
  	& &	\left. + J_1 \left(  | n, B \ket \bra n+1, A | + | n+1, A \ket \bra n, B | \right)	 \right]
				\nonumber
	,
\eeqa
We write a generic wave solution for this model in the form $| \psi \ket = \sum_{n=-\infty}^\infty \left( C_A | n, A \ket + C_B | n, B \ket \right) e^{ink}$.  Plugging this ansatz into the Schr\"odinger equation $\Hinf | \psi \ket = E |\psi \ket$ we can obtain an equivalent effective eigenvalue equation in the single-cell subspace of the model in the form
\beqa
  \left[  \begin{array}{cc}
		0				& J_2 + J_1 e^{-ik}			\\
		J_2 + J_1 e^{ik}	& 0	
		 	\end{array} \right]
	\left[  \begin{array}{c}
		C_A 	\\
		C_B 
			\end{array} \right]
	= E
	\left[  \begin{array}{c}
		C_A 	\\
		C_B 
			\end{array} \right]
		.
 \label{inf.eigen.matrix}
\eeqa
This is easily solved to yield the continuum eigenstates
\beq
   | \psi \ket = | k, \pm \ket
   	= \sum_{n=-\infty}^\infty  \frac{1}{\sqrt{2}}\big( f_k | n, A \ket \pm | n, B \ket \big) e^{ikn}
\label{inf.states}
\eeq
with the respective eigenvalues
\beq
  E_{k,\pm} = \pm \left| w_k \right|
  	\ = \ \pm \sqrt{J_1^2 + J_2^2 + 2 J_1 J_2 \cos k }
\label{E.k.pm}
\eeq
in which
\beq
  f_k = \sqrt{w_k^*/w_k}
\eeq
and
\beq
  w_k = J_2 + J_1 e^{ik}
  	.
\eeq
Equation (\ref{E.k.pm}) defines the two continuum bands, which lie on the domain $k \in \left[  -\pi, \pi \right]$.

However, the existence of two further {\it generalized} eigenstates can be revealed by setting $E=0$ in Eq. (\ref{inf.eigen.matrix}).  Doing so, we find the conditions $C_A \neq 0, C_B = 0$ yield the eigenstate
\beq
  | \psi_A \ket = \sum_{n=-\infty}^\infty \left( -1 \right)^n e^{\log \left( \frac{J_2}{J_1} \right) n} | n, A \ket 
\label{inf.psi.A}
\eeq
(after choosing $C_A = 1$).
Meanwhile the conditions $C_A = 0, C_B \neq 0$ instead give
\beq
  | \psi_B \ket = \sum_{n=-\infty}^\infty \left( -1 \right)^n e^{- \log \left( \frac{J_2}{J_1} \right) n} | n, B \ket 
\label{inf.psi.B}
\eeq
(after choosing $C_B = 1$).  Notice that $| \psi_A \ket$ grows exponentially on the left side of the chain for $J_1 > J_2$, while $| \psi_B \ket$ grows on the right side.  Then for $J_1 < J_2$ these two states switch roles. 
This demonstrates that the distinction between the topological and trivial phases is essentially lost in the infinite extension of the SSH model.  Meanwhile the sublattice localization of these two edge-like states has become exact.

However, since $| \psi_A \ket$ and $|\psi_B \ket$ are non-normalizable discrete states residing outside of the Hilbert space, they are not proper eigenstates and hence do not appear in the standard form of the diagonalized Hamiltonian \cite{HO14}.  The diagonalized form of Eq. (\ref{inf.ham}) instead consists entirely of the continuum states (\ref{inf.states}) and is given by
\beq
  H_{\textrm{SSH},\infty}
  	=  \int_{-\pi}^\pi dk \; \sum_{s = \pm} E_{k,s} | k, s \ket \bra k, s | 
	.
\label{inf.ham.diag}
\eeq

In what follows, we will use the continuum eigenstates we have obtained here to construct those for the semi-infinite SSH chain.  We will find, however, that the fate of the edge modes is quite different in the semi-infinite limit.

We emphasize that in all of these infinite-limit models, the band spectrum in Eq. (\ref{E.k.pm}) is insensitive to the boundary conditions \cite{PPT91}.  Hence, that much will remain unchanged as we move into the semi-infinite case.


\subsection{Semi-infinite SSH chain: continuum eigenstates, winding number and edge state}\label{sec:diag.semi}

We now turn to finding the eigenstates for our bath, the bare semi-infinite SSH chain, which in the single-particle picture takes the form
\beqa
  \Hbath
	& = & \sum_{n=1}^\infty \Big[ J_2 \left( | n, A \ket \bra n, B | + | n, B \ket \bra n, A | \right) 
  			\label{semi.ham}	\\
  	& &	
	+ J_1 \left(  | n, B \ket \bra n+1, A | + | n+1, A \ket \bra n, B | \right)	 \Big]
				\nonumber
	.
\eeqa
It is natural to expect that the continuum eigenstates for the semi-infinite chain might somehow be related to those from the infinite case.  
We take the simplest possible linear combination in the form of a right-moving and a left-moving wave from Eq. (\ref{inf.states}) to write
\beq
  | \phi_{k,\pm} \ket
  =    \sum_{n=1}^\infty  \left(\frac{f_k e^{ikn}  - f_k^* e^{-ikn}}{2i}  | n, A \ket 
  					\pm \sin(kn)  | n, B \ket \right)
	,
\label{semi.states}
\eeq
after projecting onto the semi-infinite subspace \footnote{We note that even in the absence of such a projection, this combination of states would satisfy $\bra 0, B | \phi_k, \pm \ket = 0$, such that if one applied the infinite Hamiltonian (\ref{inf.ham}) to these states, they would decouple from the sites to the left of $| 0, B \ket$ on the infinite chain.  It may also be instructive to compare the eigenstates (\ref{semi.states}) with those from the uniform semi-infinite tight-binding chain \cite{LonghiEPJ07}.}.  
When we apply the semi-infinite Hamiltonian (\ref{semi.ham}) to these states, we find the appropriate eigenvalue equation is indeed satisfied as
  $\Hbath | \phi_k, \pm \ket = E_{k,\pm} | \phi_k, \pm \ket$,
confirming these as the correct continuum eigenstates.

Next we consider the edge states in this semi-infinite limit.  In fact, we can address this point by applying an analysis that is conceptually similar to that typically used for the finite SSH model \cite{Asboth2016}.  In the finite case, one can predict the existence of the edge states based on the well-known bulk-boundary correspondence.  In this analysis, a winding number associated with the band eigenstates in the `bulk' Hamiltonian (the momentum-space representation of the SSH model under periodic boundary conditions) is used to predict the presence of the two edge states in the associated model after applying open boundary conditions.

\begin{figure}
 \includegraphics[width=0.45\textwidth]{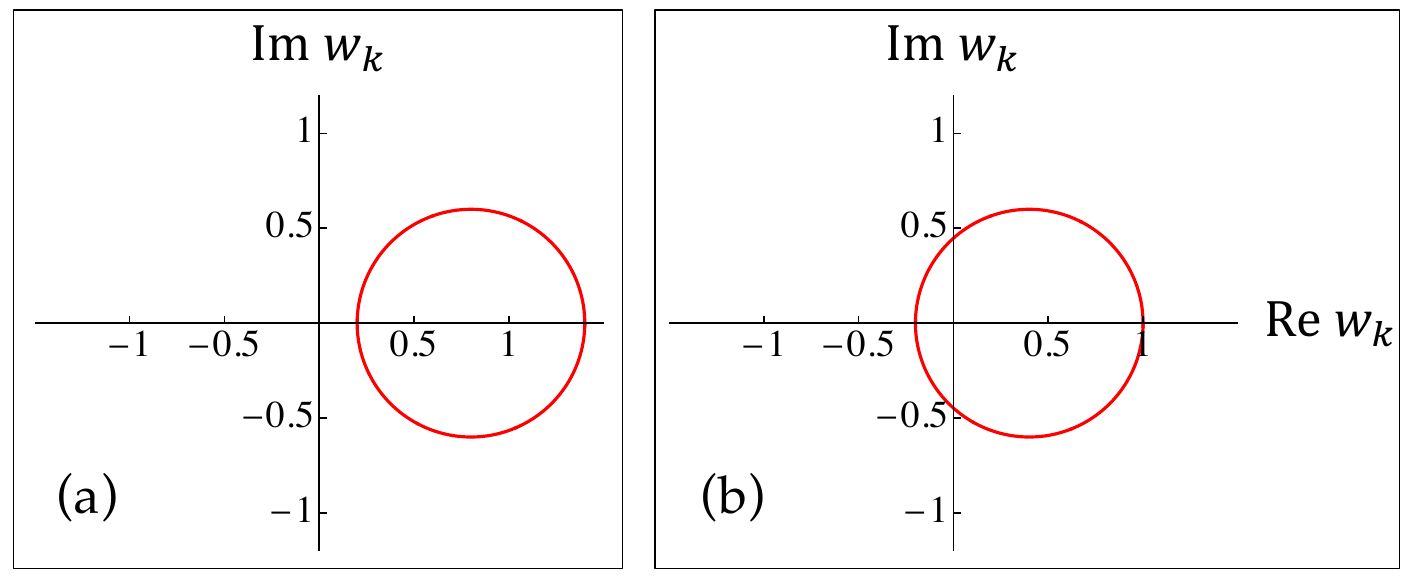}
\vspace*{\baselineskip}
\caption{Winding factor $w_k = J_2 + J_1 e^{ik}$ in the complex plane 
as $k$ runs from $-\pi$ to $\pi$ for (a) trivial case with $J_1 = 0.6$ and $J_2 = 0.8$; (b) topological case with $J_1 = 0.6$ and $J_2 = 0.4$.
}
\label{fig:semi.WN} 
\end{figure}

In the present context, we simply treat the continuum eigenstates above as the bulk states for the semi-infinite SSH model, and from them we find that we can introduce a winding number that predicts the edge state properties in the {\it same} Hamiltonian. 
Notice from Eq. (\ref{semi.states}) that the mathematical structure of the coefficient for the $A$ sites is rather different than that for the $B$ sites.  In particular, the expression $w_k^{1/2}$ appearing in the denominator of $f_k$ 
suggests the $A$-site amplitudes  might have the potential to exhibit a pole, while the $\sin (kn)$ coefficient for the $B$-site amplitudes gives no such possibility.  Hence, we introduce the winding number $\nu_A$ as the number of times $w_k =  J_2 + J_1 e^{ik}$ winds around the origin as $k$ is swept from $-\pi$ to $\pi$.  As seen in Fig. \ref{fig:semi.WN}, this winding number is zero for $J_1 < J_2$ while it becomes one for $J_1 > J_2$.  

As we demonstrate below, the following interpretation can be given.  
For the case $J_1 > J_2$ (topological case), the non-zero winding number $\nu_A = 1$ signifies that the left edge state (on the finite end of the semi-infinite chain) becomes an exact eigenstate in the semi-infinite limit, which now exhibits perfect localization onto the $A$ sublattice.  
But in the trivial case $J_1 < J_2$, this state is no longer a proper eigenstate.  (More precisely, we find below that it transitions to an anti-bound state, which is an unnormalizable state appearing in the second Riemann sheet of the complex energy plane \cite{Nussenzveig59,Hogreve,Moiseyev_NHQM,HO14}.)  Meanwhile, the fact that no winding number can be assigned to the $B$ sites at all implies that the right edge state (corresponding to the infinite side of the semi-infinite chain) has instead disappeared from the spectrum entirely. 

We can prove these statements by analysis of the resolution of identity for the present model as follows.  
First, let us hypothesize that no bound state with discrete eigenvalue exists.  Then the resolution of identity would only include the scattering states and could be written
\beq
  I_\textrm{trivial} = \int_{-\pi} ^ {\pi} \frac{dk}{2\pi} \sum_{s=\pm} |\phi_k,s\ket\bra \phi_k, s|
  	.
\label{I.trivial}
\eeq
If we insert this into the expression $\bra n, A|n', A\ket = \delta_{n,n'}$, we find
\beqa
  \bra n, A|n', A\ket  
  	&=& \delta_{n,n'} -
 		\int_{-\pi} ^ {\pi} \frac{dk}{2\pi} (f_k)^2 e^{ik(n+n')} 
\label{nA.nA.integral}			\\
	&=&\delta_{n,n'} 
		- \frac{1}{2\pi i} \oint d \lambda\, \frac{J_2 \lambda+ J_1}{J_2 + J_1 \lambda}\, \lambda^{n+n'-2}
		\nonumber
\eeqa
after changing the integration variable from $k$ to $\lambda = e^{ik}$. 
Of course, this expression must simply give $\bra n, A|n', A\ket = \delta_{n,n'}$ to be correct.  In the case $J_1 < J_2$, the winding factor  $J_2 + J_1 e^{ik}$ appearing in the denominator gives no pole inside the unit circle and hence the integral in Eq. (\ref{nA.nA.integral}) vanishes.  This confirms that the resolution of unity is given by the trivial expression in Eq. (\ref{I.trivial}) for $J_1 < J_2$, which means there is no edge state in this case.

However, in the case $J_1 > J_2$, a pole does appear in the integral in Eq. (\ref{nA.nA.integral}), which indicates the resolution of unity in Eq. (\ref{I.trivial}) is incomplete.  To cancel this non-vanishing term in Eq. (\ref{nA.nA.integral}) we find it necessary to add the missing (left) edge state $|\phi_e\ket$, corresponding to the finite end of the chain.  The resolution of unity now takes the form
\beq
  I_\textrm{top} 
  	= \int_{-\pi} ^ {\pi} \frac{dk}{2\pi} \sum_{s=\pm} |\phi_k,s\ket\bra \phi_k, s|
			+ |\phi_e\ket \bra \phi_e|
  	.
\label{I.top}
\eeq
Working out the eigenvalue problem in the site representation, it is straightforward to show that the eigenvalue associated with $|\phi_e\ket$ is $E_e = 0$ while the eigenstate itself takes the form
\beq
 |\phi_e\ket 
 	= \frac{\sqrt{J_1^2 - J_2^2}}{J_1}  \sum_{n=1}^\infty  (-)^{n-1} e^{ -(n-1)/ \xi }   |n, A \ket 
 	.
\label{semi.edge}
\eeq
in which the characteristic localization length is given by
\beq
  \xi \equiv \frac{1}{\log \left( J_1 / J_2 \right)}
  	.
\label{semi.edge.xi}
\eeq
Notice that this edge state exhibits both exponential localization (for $J_1 > J_2$) to the left side of the chain as well as sublattice localization onto the $A$ sites.  These results confirm our interpretation of the winding number $\nu_A = 1$ for $J_1 > J_2$ discussed above.  However, $|\phi_e\ket $ becomes an anti-bound state for $J_1 < J_2$, in which case $\nu_A = 0$.

Meanwhile, performing the comparable calculation for the $B$ sites reveals $\bra n, B |n', B \ket = \delta_{n,n'}$ holds regardless of the parameter values.  Hence we find that the hypothetical winding number $\nu_B$ simply vanishes.

It is an interesting peculiarity to observe that the fate of the edge states in the infinite extension of the SSH model and the semi-infinite extension have turned out to be rather different.


\section{Semi-infinite SSH bath with quantum emitter: continuum states and bound states}\label{sec:1A}

We next focus on the model in Eq. (\ref{1A.ham.2q}), in which a quantum emitter is coupled to the semi-infinite SSH chain that now acts as a reservoir.  First, we note this model also exhibits a form of chiral symmetry $\bar{\Sigma}_C H \bar{\Sigma}_C = - H$, similar to the original SSH model.  The chiral operator
$\bar{\Sigma}_C = \hat{P}_A - \hat{P}_{BQ}$ is almost the same as that from the SSH model, except that the projection operators $\hat{P}_A = \sum_{n=1}^\infty | n, A \ket \bra n, A | $ and  $\hat{P}_{BQ} = | q \ket + \sum_{n=1}^\infty | n, B \ket \bra n, B |$ now extend to infinity and $\hat{P}_{BQ}$  includes the quantum emitter $| q \ket$ as an effective $B$ site.   Hence, the eigenstates again come in chiral pairs with opposite eigenvalues $E$ and $-E$, except for exact zero-energy modes, which are self-chiral.

In the following, we will obtain the new continuum eigenstates in the site representation by accounting for the modified boundary condition after incorporating the quantum emitter in Eq. (\ref{1A.ham.2q}).  From this, we will see how the winding number picture is modified from the bare SSH chain, which enables us to identify all bound states in the model.  

Since the emitter coupled to the endpoint influences the boundary condition on the left side of the chain, we expect the eigenstates will be modified from those of the bare SSH chain in Eq. (\ref{semi.states}) primarily by a change in the amplitudes for the sites directly connected with the emitter as well as an overall phase.  Hence, we write our ansatz
\beqa
  |\psi_k, \pm \ket
  	&=&   \sum_{n=2}^\infty   |n,a\ket \frac{ \beta_k  f_k e^{ikn}  - \beta_{-k} f_{-k} e^{-ikn}}{2i}  \nonumber \\
	&  &	\pm \sum_{n=1}^\infty |n,b\ket \frac{ \beta_k  e^{ikn}  - \beta_{-k} e^{-ikn}}{2i}   \nonumber \\
  	&+& |1,A\ket\bra 1,A |\psi_k,\pm\ket  + |q\ket\bra q |\psi_k,\pm\ket 
  \label{1A.states.gen}
\eeqa
with the undetermined amplitudes $\bra q |\psi_k,\pm\ket$ and $\bra 1,A |\psi_k,\pm\ket$ as well as the phase $\beta_k$.
 We can apply the eigenvalue equation
$H |\psi_k, \pm\ket = E_{k,\pm} |\psi_k, \pm\ket$ to obtain useful relations between these unknown quantities.  
Eliminating $\bra q| \psi_k,\pm\ket $ and $\bra 1, A | \psi_k,\pm\ket $ from the resulting expressions in favor of 
$\bra 1, B | \psi_k,\pm\ket $ and $\bra 2, A |\psi_k,\pm\ket$
we eventually obtain
\beqa
\bra 1,B  |\psi_k,\pm\ket  
	= \pm \frac{J_1}{E_{k,+} -\frac{J_2^2}{E_{k,+} -g^2/E_{k,+}}}  \bra 2, A |\psi_k,\pm\ket
	,
\eeqa
where we have replaced $E_{k,\pm}$ with $\pm E_{k,+}$ and factored out an overall sign.
This gives the new boundary condition of the bulk states, and, after applying Eq. (\ref{1A.states.gen}) again,
yields 
an equation for $\beta_k$ that can be solved as \footnote{Note that  $\beta_k \to1$ when $g\to 0$, so that when the emitter is removed the continuum states in Eq. (\ref{1A.states.gen}) correctly reduce to the continuum states from the bare semi-infinite SSH chain in Eq. (\ref{semi.states})}
\beq
\beta_k = \sqrt{\frac{w_k {\tilde w}_{-k}}{w_{-k} {\tilde w}_{k}}}
\label{1A.betak}
\eeq
where
\beq
{\tilde w}_{k} = J_2 + {\tilde J}_1 e^{i k}, \quad {\tilde J}_1 = J_1 - g^2/J_1
	.
\eeq
Applying Eq. (\ref{1A.betak}) in (\ref{1A.states.gen}), the continuum eigenstates finally take the form
\begin{widetext}
\beqa
|\psi_k, \pm \ket
  &=&   \frac{1}{2 i} \sum_{n=2}^\infty   |n,A\ket \left(\sqrt{\frac{{\tilde w}_{-k}}{{\tilde w}_{k}}} e^{ikn}  - \sqrt{\frac{{\tilde w}_{k}}{{\tilde w}_{-k}}} e^{-ikn} \right)
  \pm \frac{1}{2 i} \sum_{n=1}^\infty |n,B\ket  \left( \sqrt{\frac{w_k {\tilde w}_{-k}}{w_{-k} {\tilde w}_{k}}}e^{ikn}  -  \sqrt{\frac{w_{-k} {\tilde w}_{k}}{w_{k} {\tilde w}_{-k}}} e^{-ikn}\right)   \nonumber\\
  &+& |1,A\ket\bra 1,A |\psi_k,\pm\ket  + |q\ket\bra q |\psi_k,\pm\ket
  .
  \label{1A.states}
\eeqa
\end{widetext}
By comparing with Eq. (\ref{semi.states}) for the bare semi-infinite chain, we observe that the factor $f_k = \sqrt{w_{-k}/w_k}$ no longer appears in the coefficient for the $A$ sites; instead, the reciprocal of this expression now appears before the $B$ sites.  This immediately tells us that the winding number $\nu_A$ vanishes once we attach the quantum emitter at the endpoint and, further, that now $\nu_B$ is the non-trivial winding number.  But since $f_k$ is inverted in Eq. (\ref{1A.states}), we find that the phases are also inverted, such that now $J_1 < J_2$ gives $\nu_B = 1$ while $J_1 > J_2$ gives $\nu_B = 0$.  
Hence, we expect to find a zero-energy bound state $| \psi_0 \ket$ localized on the $B$ sublattice for $J_1 < J_2$.  
Solving the eigenvalue equation $H | \psi_0 \ket = 0$ we indeed find the expected state
\beq
  | \psi_0 \ket = \bra q | \psi_0 \ket \left[ | q \ket + \frac{g}{J_2} 
  		\sum_{n=1}^\infty \left( -1 \right)^n \left( \frac{J_1}{J_2} \right)^{n-1} | n, B \ket \right]
\label{psi.zero}
\eeq
with eigenvalue $z_0 = 0$ and normalization
\beq
   \bra q | \psi_0 \ket = \sqrt{ \frac{J_2^2 - J_1^2}{J_2^2 - J_1^2 + g^2} }
   	.
\label{psi.zero.norm}
\eeq

Most interestingly, however, we see that an additional overall phase ${\tilde w}_{k} / {\tilde w}_{-k}$ now appears on {\em both} the $A$ site and $B$ site terms in Eq. (\ref{1A.states}).  This introduces a new winding number that we label $\mu_{AB}$, which requires a slightly more complex interpretation.  One might guess that $\mu_{AB} \neq 0$ would imply the coexistence of two states, with one appearing on the $A$ sublattice and the other on the $B$ sublattice.  While it is ultimately possible to construct two such states, they are not eigenstates of the Hamiltonian in Eq. (\ref{1A.ham.2q}). Instead, the actual eigenstates are two linear combinations of these sublattice-localized states, which we will report momentarily.  
First, we simply use the factor $ {\tilde w}_{k}$ to predict the presence or absence of these states.

Analysis from the winding factor $ {\tilde w}_{k}$ is rather similar to that carried out previously for the bare semi-infinite chain in Sec. \ref{sec:diag.semi} and reveals the winding number $\mu_{AB} \neq 0$ in {\it three} distinct parameter regions.  First, let us consider the case $J_1 < J_2$, corresponding to the trivial phase of the bare SSH reservoir.  
In this case, we find that $\mu_{AB} = 1$ if we impose a further condition on the emitter-chain coupling $g$ as
\beq
  g >  g_\textrm{str} \equiv  \sqrt{J_1 (J_1 + J_2)}
  	.
\label{g.strong}
\eeq
This yields a pair of bound states with energy eigenvalue given by
\beq
  E_\pm = \pm g \sqrt{\frac{g^2 - J_1^2 + J_2^2}{g^2 - J_1^2}}
  	,
\label{E.pm}
\eeq
which 
appear {\it outside} of the two continuum bands
  $\left| E_\pm \right| > J_1 + J_2$.
These are examples of the ordinary strong-coupling bound states discussed in the introduction.  

However, in the case $J_1 > J_2$ (corresponding to the non-trivial phase of the bare SSH reservoir), we find that $\mu_{AB} = 1$ in two disconnected parameter regions.  The first of these is again given by the strong-coupling condition $g > \gstr$ in Eq. (\ref{g.strong}), which yields two bound states that are qualitatively similar to the case above.
But the second parameter region for $J_1 > J_2$ is instead given by the weak-coupling condition 
\beq
  g < \gw \equiv \sqrt{J_1 (J_1 - J_2)}
  	.
\label{g.weak}
\eeq
The bound state eigenvalues in this case are again given by Eq. (\ref{E.pm}), but these now appear {\it inside} the SSH band gap $\left| E_\pm \right| < J_1 - J_2$.

\begin{figure}
 \includegraphics[width=0.45\textwidth]{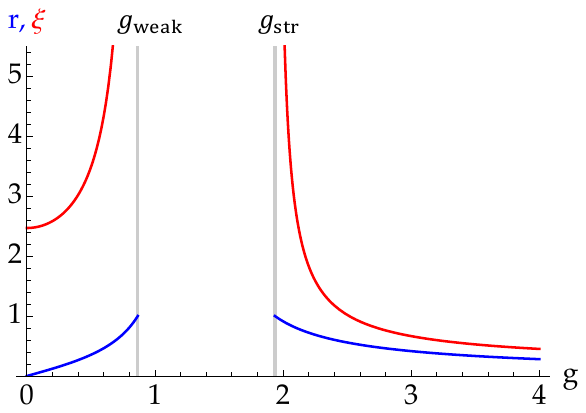}
\hfill
\\
\caption{ Localization measures -- sublattice localization $r$ (blue curve) and exponential localization length $\xi$ (red curve) as a function of $g$ for the case $J_1 > J_2$ with  $J_1 = 1.5, J_2 = 1$.  The values of 
$\gw \equiv \sqrt{J_1 (J_1 - J_2)}$ and $\gstr \equiv \sqrt{J_1 (J_1 + J_2)}$ are indicated by vertical lines.
}
 \label{fig:1A.local}
 \end{figure}

Two further qualitative distinctions can be made between the strong-coupling and weak-coupling bound states. 
The first of these is that because the weak-coupling bound states are located inside the band gap, they experience partial sublattice localization (similar to the edge states in the original, finite SSH model).  This can be illustrated from the explicit form of the eigenstates,
which are obtained in App. \ref{app:bs} in the form
\begin{widetext}
\beqa
  | \psiBpm \ket 
 	=  \bra q | \psiBpm \ket \left[ | q \ket 
		+ \frac{| E_\pm | }{g} \sum_{n=1}^\infty 
				\left( - \sign (\tilde{J}_1) \right)^{n-1} \Big( \pm | n,A \ket - r | n, B \ket \Big) e^{-(n-1)/\xi}  \right]
\label{psi.BS}
\eeqa
\end{widetext}
with the normalization $\bra q | \psiBpm \ket$ given in the appendix Eq. (\ref{app2_6})
and in which the exponential localization length is given by
\beq
  \xi \equiv \frac{1}{\log \left| \tilde{J}_1 / J_2 \right| } 
  	.
\label{psi.BS.xi}
\eeq
Further, the factor $r$ appearing before the $| n , B \ket$ component in Eq. (\ref{psi.BS}) is given by
\beq
	r = \pm \frac{g J_2}{\sqrt{(J_1^2 - J_2^2 - g^2)(J_1^2-g^2)}}\qquad {\rm for}\, J_1\gtrless g
	.
\label{psi.BS.r}
\eeq
Equation (\ref{psi.BS}) gives the eigenstates for the pair of bound states in all three of the parameter regions described above; however, we can see that only the weak-couping bound states in the case $g < \gw$ and $J_1 > J_2$ experience partial sublattice localization as follows.
Notice that the relative magnitude of the coefficient for the $| n, A \ket$ component in Eq. (\ref{psi.BS}) is simply unity, while that for the $| n, B \ket$ component is $\left| r \right|$.  Hence, the quantity $r$ provides a natural measure of the degree of sublattice localization. 
We plot $r$ (blue curve) as a function of $g$ in the case $J_1 > J_2$ ($J_1 = 1.5, J_2 = 1$) in Fig. \ref{fig:1A.local}.  While $r = 1$ exactly at $g = \gw$, we see that as $g$ decreases so that $g \ll \gw$, we have $r \ll 1$, showing that the two eigenstates in Eq. (\ref{psi.BS}) mostly localize onto the $A$ sublattice (notice that the two bound state wave functions have opposite sign on this sublattice).  
We can also see this point analytically from the form of Eq. (\ref{psi.BS.r}) by noting that $g \ll \gw$ and $J_1 < J_2$ imply $g \ll J_1$, which yields $r \sim g \ll 1$ for the weak-coupling bound states.
Meanwhile, for the strong-coupling bound states with $g > \gstr$, we see in Fig. (\ref{fig:1A.local}) that $r$ is around unity for $g \gtrsim \gstr $, yielding parity between the two sublattices.  
From the form of Eq. (\ref{psi.BS.r}), one might think that the bound states again become sublattice localized for very large $g \gg \gstr$.  However, in this case, the exponential localization length quickly becomes $\xi \lesssim 1$, which indicates that, actually, the $| q \ket$ and $| 1, A \ket$ sites form a dimerized pair in this limit.  This can be seen from the red curve (for $\xi$) in Fig. \ref{fig:1A.local}.

\begin{widetext}

\begin{figure}
\hspace*{0.05\textwidth}
 \includegraphics[width=0.4\textwidth]{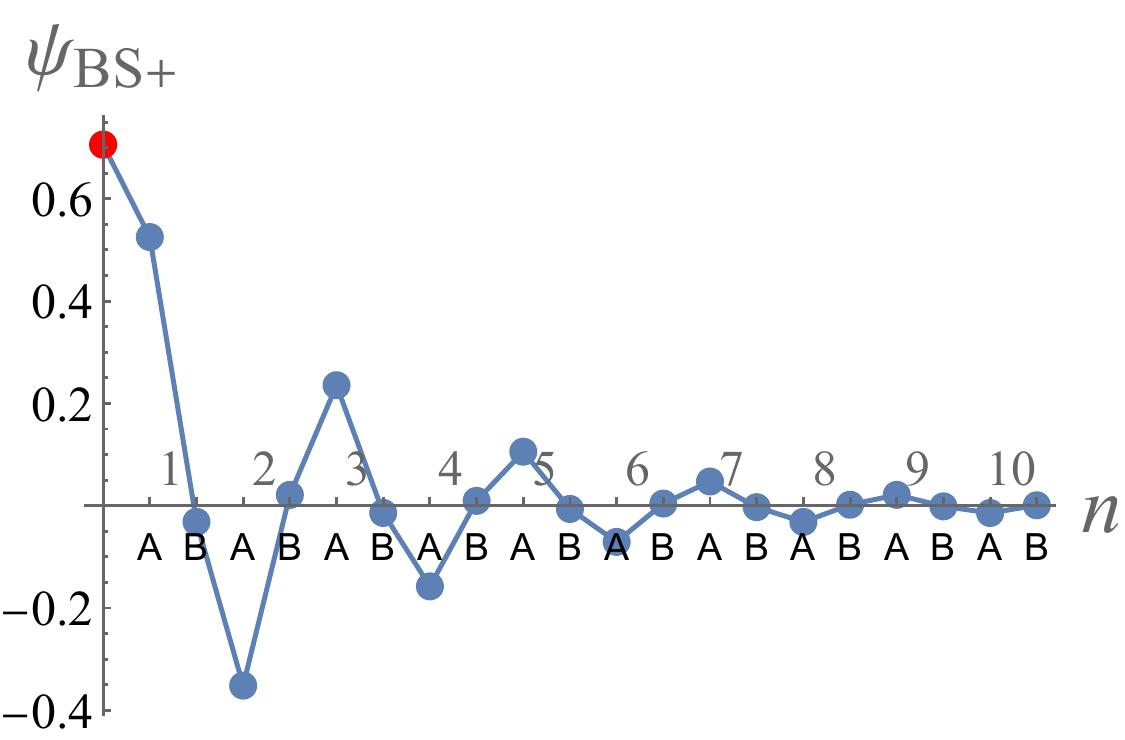}
\hfill
 \includegraphics[width=0.4\textwidth]{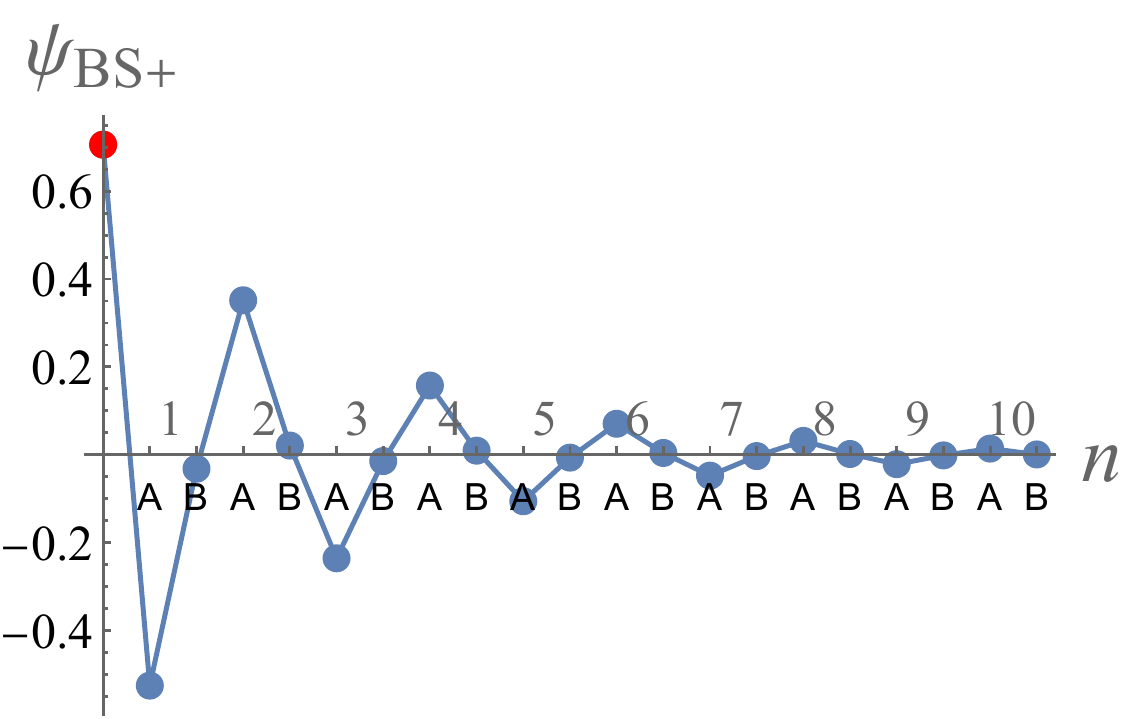}
 \hspace*{0.05\textwidth}
\\
\vspace*{\baselineskip}
\hspace*{0.05\textwidth}(a)\hspace*{0.440\textwidth}(b)\hspace*{0.4\textwidth}
\\
\vspace*{\baselineskip}
\hspace*{0.05\textwidth}
 \includegraphics[width=0.4\textwidth]{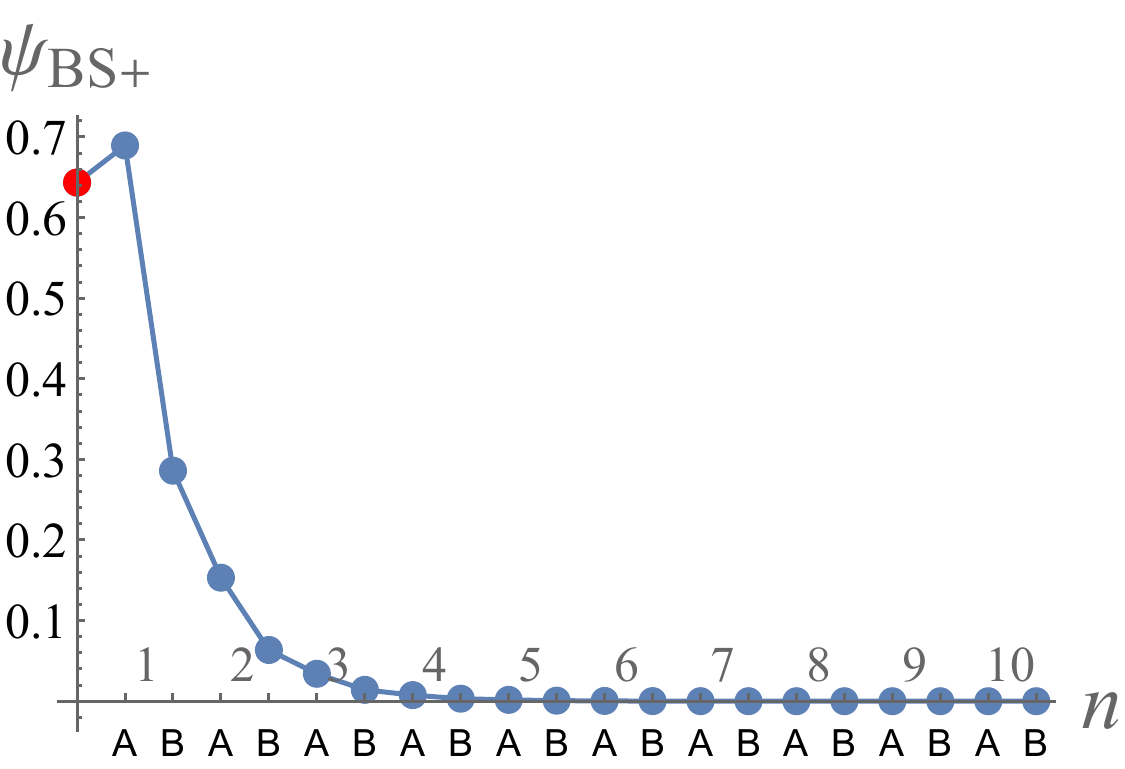}
\hfill
 \includegraphics[width=0.4\textwidth]{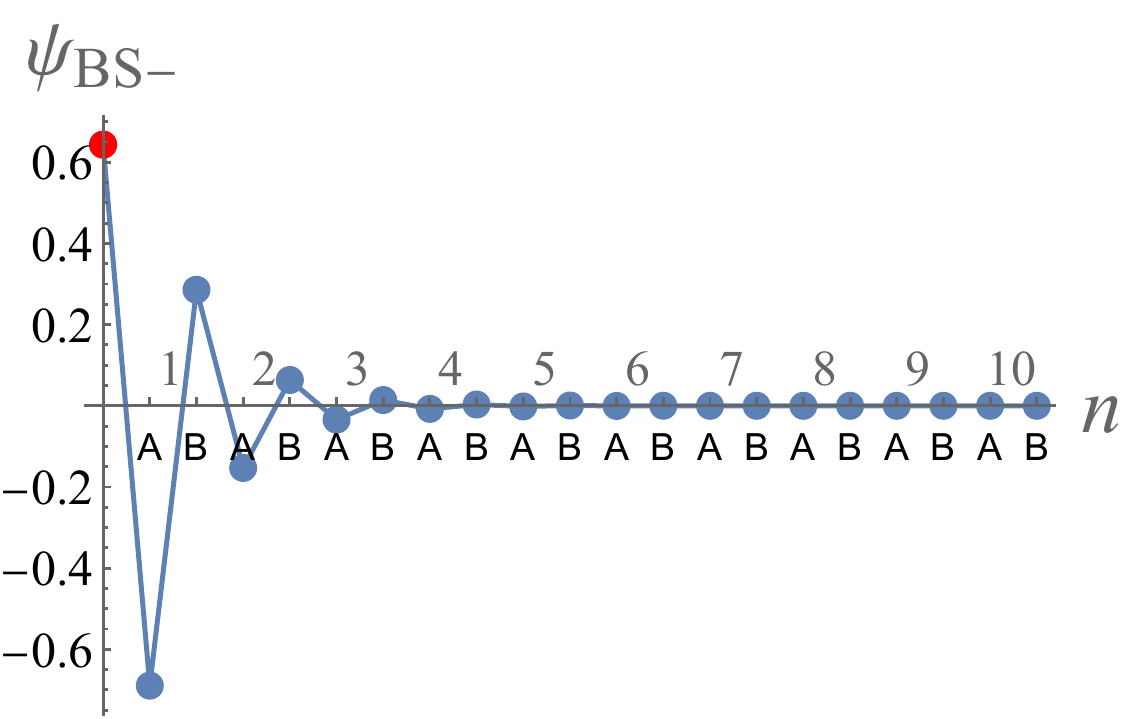}
 \hspace*{0.05\textwidth}
\\
\vspace*{\baselineskip}
\hspace*{0.05\textwidth}(c)\hspace*{0.440\textwidth}(d)\hspace*{0.4\textwidth}
\\
\vspace*{\baselineskip}
\caption{Wave function for the (two) bound states in two representative cases: (a, b) weak-coupling bound states in the case $g=0.1$, and (c, d) strong-coupling bound states in the case $g=3.0$.  We use $J_1 = 1.5, J_2 =1.0$ in all figures.  The red dot indicates the amplitude of the wave function at the quantum emitter.
}
 \label{fig:bound.wf}
 \label{fig7}
 \end{figure}

\end{widetext}

The third qualitative distinction between the weak-coupling and strong-coupling bound states can be seen from the factor $( - \sign (\tilde{J}_1) )^{n-1}$ in Eq. (\ref{psi.BS}).  This factor implies that the weak-coupling bound states (for which $\sign (\tilde{J}_1) = +$) experience an alternating sign between the amplitudes on the $n^{\rm th}$ and $(n+1)^{\rm th}$ cells, which again is similar to the behavior of the edge states in the finite SSH model.  
Meanwhile, this alternating sign is canceled out for the strong-coupling bound states (instead, the strong-coupling bound state with negative energy experiences a sign change between $A$ and $B$ sites within each cell).  This can be seen for representative cases in Fig. \ref{fig:bound.wf}.

The reason that the weak-coupling bound states are qualitatively similar to edge states is easy to understand if we consider the small-coupling limit.  Assuming $g$ is non-zero but satisfies  $g \ll \gw$, a simple perturbative calculation can be performed to show that the two weak-coupling bound states are approximately two hybridizations between the quantum emitter and the edge state of the bare semi-infinite SSH chain, taking the form
\beq
  | \psiBpm \ket \approx \frac{1}{\sqrt{2}} \big( | q \ket \pm | \phi_e \ket \big)
  	.
\label{psi.BS.hybrid}
\eeq
This also shows that the weak-coupling bound states can be viewed as a line-splitting phenomenon that occurs within the topologically-induced band gap.

Returning to our discussion above about the winding number, notice from Eq. (\ref{psi.BS}) that we can easily construct two states that are localized entirely on either the $B$ or $A$ sublattice by writing
\beq
  | \psi_{s,a} \ket \equiv \frac{1}{\sqrt{2}} \Big( | \psiBp \ket \pm | \psiBm \ket \Big)
\label{psi.sa}
\eeq
with $| \psi_{s} \ket$ ($| \psi_{a} \ket$) residing exclusively on the $B$ ($A$) sublattice.   
We note these new states are somewhat like the edge states in the original SSH model, except that they are both localized to the left side of the chain.  


The $| \psi_{s,a} \ket$ states further have the following interesting properties.
First, they transform into each other under the action of the Hamiltonian, according to
  $H | \psi_{s,a} \ket = E_+ | \psi_{a,s} \ket
  	.$
While this illustrates that $ | \psi_{s,a} \ket$ are not eigenstates of $H$, they clearly are eigenstates of $H^2$, as
\beq
  H^2 | \psi_{s,a} \ket = E_+^2 | \psi_{s,a} \ket
  	.
\eeq
This in turn implies that these states follow a relatively simple dynamical evolution (in a closed subspace) according to
\beq
  e^{-i H t}  | \psi_{s,a} \ket 
	= \cos E_+ t  \ | \psi_{s,a} \ket  - i \sin E_+ t \ | \psi_{a,s} \ket
		.
\label{psi.sa.time}
\eeq
This relationship will turn out to have interesting implications when we consider the time evolution of the quantum emitter in the weak-coupling scenario later, in Sec. \ref{sec:1A.dynamics.topological}.


\section{Generalized spectrum from the Green's function}\label{sec:1A.spectrum}

We have found in the previous section that the bound states exist in three (partly disconnected) parameter regions.  In the following, we aim to fill out this picture by obtaining the generalized spectrum for the full parameter space, which additionally includes resonance states, anti-resonance states and anti-bound states.  We emphasize that only the original bound states are exponentially-localized solutions that reside in the usual Hilbert space, while the others are non-normalizable, non-standard solutions.  Nevertheless, 
the elements of the generalized spectrum have the potential to influence the dynamics in a variety of scenarios \cite{PPT91,BCP10,HO14,GNOS19,GarmonPRR21,GPSS13}.

The generalized spectrum can be obtained from the poles of the Green's function of the quantum emitter, written in the form $\bra q | (z - H)^{-1} | q \ket $.  As shown in App. \ref{app:1A.green} this can be evaluated by a repeating fraction method to obtain
\beq
  \bra q | \frac{1}{z - H} | q \ket 
	= \frac{1}{z - \Sigma (z)}
\label{1A.q.q}
\eeq
in which the self-energy function $\Sigma (z)$ contains information about the influence of the reservoir on the particle at the quantum emitter site and is given by
\beq
  \Sigma (z) 
  = g^2 \left( \frac{z^2 + J_+ J_- + S (z)}{2 J_1^2 z} \right)
	,
\label{1A.self}
\eeq
in which
\beq
  J_\pm \equiv J_1 \pm J_2
\eeq
and
\begin{widetext}
\beq
  S(z + i 0^\pm) = \left\{  \begin{array}{cl}
  		+ \sqrt{\left( z^2 - J_+^2 \right) \left( z^2 - J_-^2 \right)}	
							&	\textrm{for} \left| z \right| > J_+	 \textrm{\ \ \ (outside the bands)} \\
		\pm i \sqrt{\left( J_+^2 - z^2 \right)  \left( z^2 - J_-^2 \right)}	
							& \textrm{for} \ J_- < \re \ z < J_+, \im \ z = 0^\pm		\\
		\mp i \sqrt{\left( J_+^2 - z^2 \right)  \left( z^2 - J_-^2 \right)}	
							& \textrm{for} \  -J_+ < \re \ z < -J_-, \im \ z = 0^\pm		\\
		- \sqrt{\left( z^2 - J_+^2 \right) \left( z^2 - J_-^2 \right)}	
							&	\textrm{for} \left| z \right| < J_-  \textrm{\ \ \ (inner band gap)}
		\end{array}
	 \right.
	 .
\eeq
\end{widetext}
With this expression in hand, we find that the eigenvalue equation $z - \Sigma (z) = 0$ from the
pole of the Green's function (\ref{1A.q.q}) yields three solutions.  The first of these is of course the zero-mode we found in Eq. (\ref{psi.zero}), which, as we previously discussed, gives a bound state for $J_1 < J_2$ or an anti-bound state in the case $J_1 > J_2$.  
The other two solutions are generalizations of the pair of bound states we detailed in Sec. \ref{sec:1A}, with the (in general) non-zero eigenvalues given by
\beq
  z_\pm = g \sqrt{\frac{g^2 - \left( J_1^2 - J_2^2 \right)}{g^2 - J_1^2}}
  	.
\label{z.pm}
\eeq
These are of course just the same as the eigenvalues for the bound states $E_\pm$ from Eq. (\ref{E.pm}), but we now use the notation $z$ for energy to denote the possibility that the eigenvalues may become complex-valued.  The corresponding eigenstates are simply given by generalization of Eq. (\ref{psi.BS}), taking the form
\begin{widetext}
\beqa
  | \psi_\pm \ket 
 	=  \bra q | \psi_\pm \ket \left[ | q \ket 
		+ \frac{| z_\pm | }{g} \sum_{n=1}^\infty 
				 \Big( \pm | n,A \ket - r | n, B \ket \Big) e^{-ik_\pm (n-1)}  \right]
\label{psi.pm}
\eeqa
\end{widetext}
in which $k_\pm$ is the shared wave vector
\beq
  k_+ = k_- = - i \log \left( \frac{J_1 J_2}{ g^2 - J_1^2 } \right) 
  	.
\label{k.pm}
\eeq
For the three parameter regions discussed in Sec. \ref{sec:1A}, we find $\im \; k_\pm > 0$, such that the wave function (\ref{psi.pm}) becomes localized and we recover the expression for the bound states in Eq. (\ref{psi.BS}).  However, in other parameter regions, we find instead that $\im \; k_\pm < 0$, such that the eigenstates (\ref{psi.pm}) are delocalized, yielding generalized eigenstates.

\begin{figure}
 \includegraphics[width=0.4\textwidth]{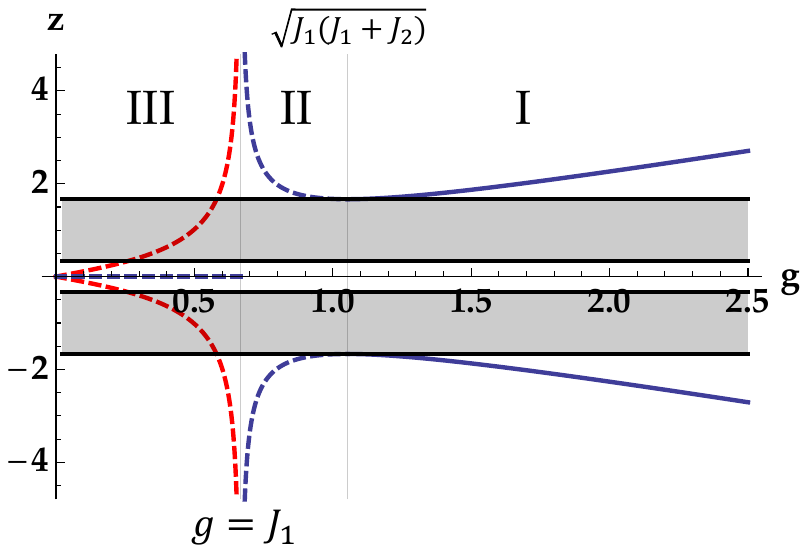}
\vspace*{\baselineskip}
\hspace*{0.01\textwidth}(a)\hspace*{0.47\textwidth}
\\
 \includegraphics[width=0.4\textwidth]{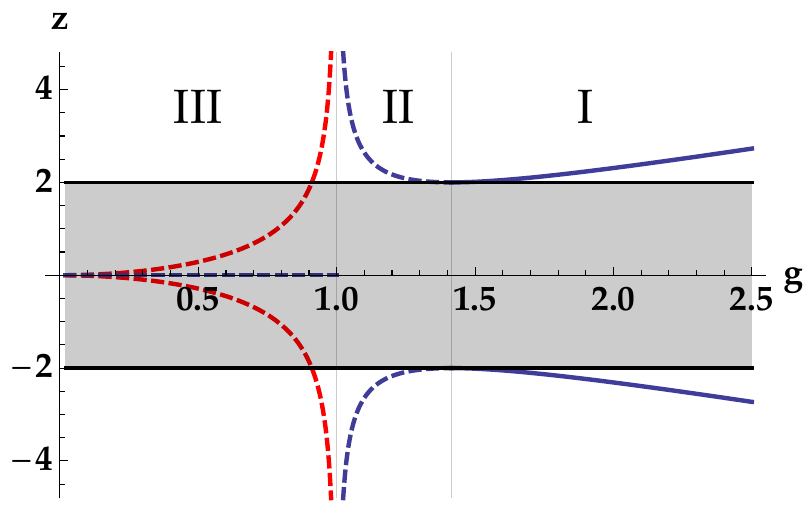}
\\
\vspace*{\baselineskip}
\hspace*{0.01\textwidth}(b)\hspace*{0.47\textwidth}
\vspace*{\baselineskip}
\caption{ (a) Energy spectrum of the system as the parameter $g$ varies, including the (purely real) continuum eigenvalues $E_{k\pm}$ and the real parts (blue line) and imaginary parts (red line) of the discrete eigenvalues $z_\pm$ for the trivial case $J_1 < J_2$ with $J_1 = 2/3$ and $J_2 = 1$.  The character of the eigenvalues $z_\pm$ change within the three regions I--III as described in the main text.  The solutions indicated with full lines reside within the first Riemann sheet, while the broken lines are within the second sheet.
The zero eigenvalue $z_0$ (bound state) is not explicitly shown.
(b) The same plot for the case of the uniform chain with $J_1 = J_2 = 1$.  There is no zero-energy eigenvalue for the uniform chain.
}
 \label{fig:1A.spec.triv}
 \end{figure}

First, let us examine the generalized spectrum for $J_1 < J_2$, corresponding to the trivial configuration of the bare SSH reservoir, which is shown for the case $J_1 = 2/3$ and $J_2 = 1$ in Fig. \ref{fig:1A.spec.triv}(a).
The two continuum bands $E_{k,\pm}$ are indicated by the two shaded regions, lying between $\pm \left( J_1 + J_2 \right)$ and $\pm \left( J_2 - J_1 \right)$.  
Meanwhile the two non-zero solutions $z_\pm \neq 0$ are shown by the curves, for which the full curve indicates a localized (bound state) solution with $\im k > 0$, while the dashed curves represent the generalized states with $\im k < 0$.  We find there are three distinct parametric regions, indicated as Regions I, II and III in Fig.  \ref{fig:1A.spec.triv}(a).  Region I, given by $g > \gstr$, corresponds to the strong-coupling bound states, with the eigenvalues appearing outside the SSH bands, as expected.  As the value of $g$ is decreased, we see the two bound state eigenvalues approach the respective nearby band edges.  At exactly $g = \gstr$, the bound state eigenvalues touch their respective band edges, as the bound state is absorbed into the continuum \cite{GPSS13}.  Then in Region II, defined by $J_1 < g < \gstr$, these two solutions become anti-bound states.  We emphasize that while these are not proper eigenstates in the usual sense, it can be demonstrated that they have a clear, predictable influence on the non-Markovian dynamics associated with the branch-point effect \cite{GPSS13,GNOS19}.  Finally, in Region III ($g < J_1$), the two anti-bound states transition to a resonance/anti-resonance pair.  The imaginary part of the eigenvalues $\im z$ is indicated in this case with a red, dashed curve.  We note that while the presence of a resonance is usually associated with exponential decay, as we show below, as long as the SSH gap is wide, the resonance does not influence the dynamics so strongly.  Finally, the localized zero mode coexists with these solutions in each region, but is not explicitly shown (to avoid overlapping with the real part of the resonance).

We can gain perspective on the spectral plot in Fig.  \ref{fig:1A.spec.triv}(a) by considering the limit $J_1 \to J_2$ in which the dimerization is replaced by a uniform chain, shown in Fig.  \ref{fig:1A.spec.triv}(b).  The spectrum of the uniform chain is qualitatively rather similar to that from the $J_2 > J_1$ case, with the same three regions I, II and III.  The primary differences are just that in the case of the uniform chain, the band gap has closed and the zero mode has vanished.

\begin{figure}
 \includegraphics[width=0.45\textwidth]{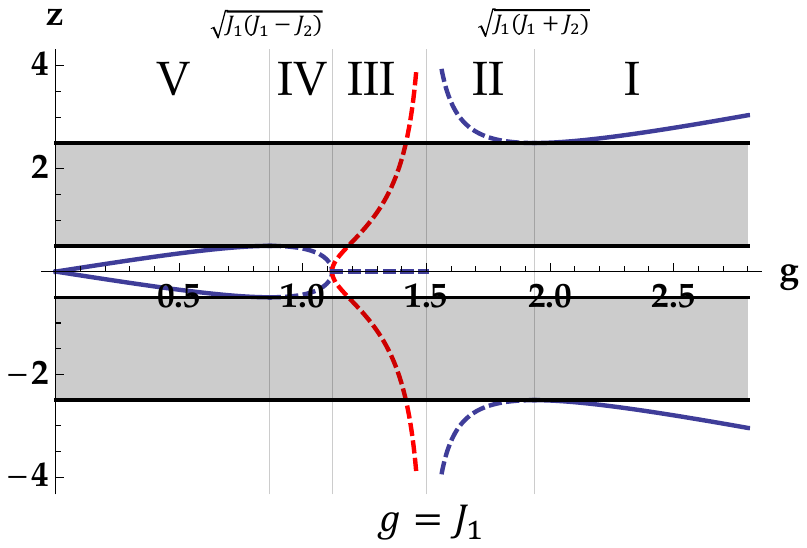}
\hfill
\\
\caption{ Energy spectrum of the system as the parameter $g$ varies, including the (purely real) continuum eigenvalues $E_{k\pm}$ and the real (blue) and imaginary (red) parts of the discrete eigenvalues $z_\pm$ for the non-trivial case $J_1 > J_2$ with $J_1 = 3/2$ and $J_2 = 1$.  The character of the eigenvalues $z_\pm$ change within the five regions I--V as described in the main text.  The solutions indicated with full lines reside within the first Riemann sheet, while the broken lines are within the second sheet.
The zero eigenvalue $z_0$ (anti-bound state) is not explicitly shown.}
 \label{fig:1A.spec.top}
 \end{figure}

However, when we consider the $J_1 > J_2$ case in Fig.  \ref{fig:1A.spec.top}, we immediately observe a non-trivial correspondence with the uniform chain.  While regions I, II and III appear as before, there are two new regions, IV and V, in which the two non-zero solutions are real-valued and appear inside the SSH band gap.  In Region IV, these solutions form two inner-gap anti-bound states, while in Region V they form the weak-coupling bound states we studied previously in Sec. \ref{sec:1A}.  Indeed, in the limit $g \to 0$, we see that these two solutions converge to a common eigenvalue at $z=0$.  Stated differently, for small coupling $g \ll \gw = \sqrt{J_1 (J_1 - J_2)}$, these two solutions split from a common eigenvalue $z = 0$, which is just the line splitting previously expressed in Eq. (\ref{psi.BS.hybrid}).  Of course, another difference from the $J_1 < J_2$ case in 
Fig.  \ref{fig:1A.spec.triv}(a) is that for $J_1 > J_2$ case the zero mode has become antibound.

Finally, we briefly note that the boundary between Regions III and IV is demarcated by an exceptional point  (EP) occurring at $g = \sqrt{J_1^2 - J_2^2}$  \cite{BerryEP,HeissEP,GGH15,EPReview}.  At this EP, the two states with non-zero eigenvalue coalesce with the zero-energy mode, qualitatively similar to several of the EPs reported in Ref. \cite{GN21}.


\section{Quantum emitter dynamics}\label{sec:1A.dynamics}

\begin{figure}
 \includegraphics[width=0.45\textwidth]{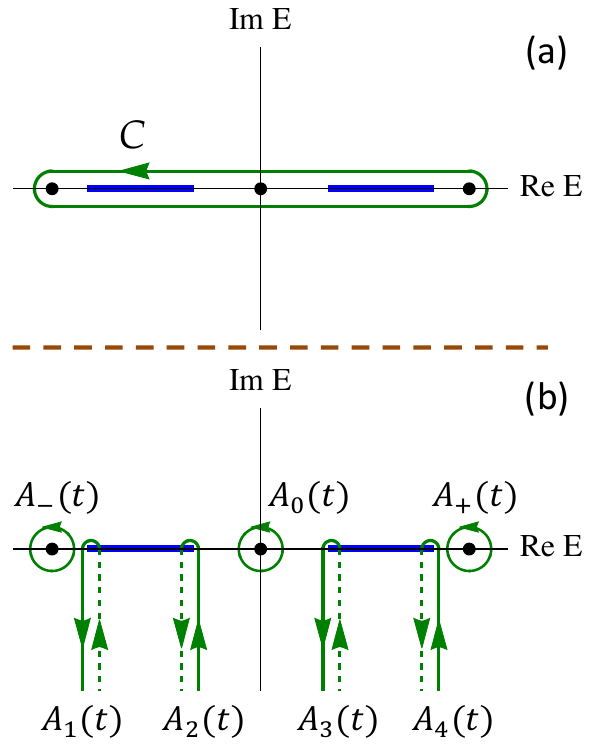}
\hfill
\\
\caption{ (a) Original integration contour $C$ (in Region I), (b) deformed contour, including the three poles associated with the bound states in Region I and all branch-point contributions, as described in the main text and App. \ref{app:1A.BPE}.  Note that full lines represent contour contributions in the first Riemann sheet while dotted lines represent contributions lying in the second sheet.
 }
 \label{fig:contour}
 \end{figure}

In this section, we primarily aim to illustrate the dynamical influence of the various bound states.
Let us assume that the particle is initialized in the quantum emitter site with absolute probability at an initial time $t=0$.  The probability that the particle remains at the quantum emitter site at some later time $t$ is determined by the survival probability $P(t) = \left| A(t) \right|^2$, in which $A(t)$ is the survival amplitude
\beqa
  A(t) = \bra q | e^{-iHt} | q \ket
	.
\label{amp.defn}
\eeqa
Apart from a very brief, early time period 
in which parabolic dynamics tend to dominate the evolution \cite{Sudarshan77}, it is usually easiest to evaluate the survival probability by rewriting it in terms of an integration over the Green's function in the form
\beqa
  A(t) 
  	= \frac{1}{2 \pi i} \int_C dz e^{-i z t}  \bra q | \frac{1}{z - H} | q \ket
	.
\label{amp.green}
\eeqa
The contour $C$ here 
encircles the entire real axis of the complex energy plane, including any bound states that are present 
as well as both energy continuum bands $E_{k,\pm}$.  This contour is shown in Fig. \ref{fig:contour}(a) for the example of Region I in the case $J_1 < J_2$.   

In what follows, we first consider the evolution in the case $J_1 < J_2$, corresponding to the trivial configuration of the bare semi-infinite SSH chain.  Then we evaluate the dynamics influenced by the weak-coupling bound states in Region V for the $J_1 > J_2$ case in Sec. \ref{sec:1A.dynamics.topological}.

\subsection{$J_1 < J_2$ case: Dynamics in Regions I-III}\label{sec:1A.dynamics.trivial}

We briefly consider the dynamics for the $J_1 < J_2$ case here, in which the SSH reservoir is in the trivial configuration.  The spectrum for this case was shown previously in Fig. \ref{fig:1A.spec.triv}(a).  We remind the reader that the zero mode takes the form of a bound state for all three parameter regions in this case.

\begin{figure}
 \includegraphics[width=0.45\textwidth]{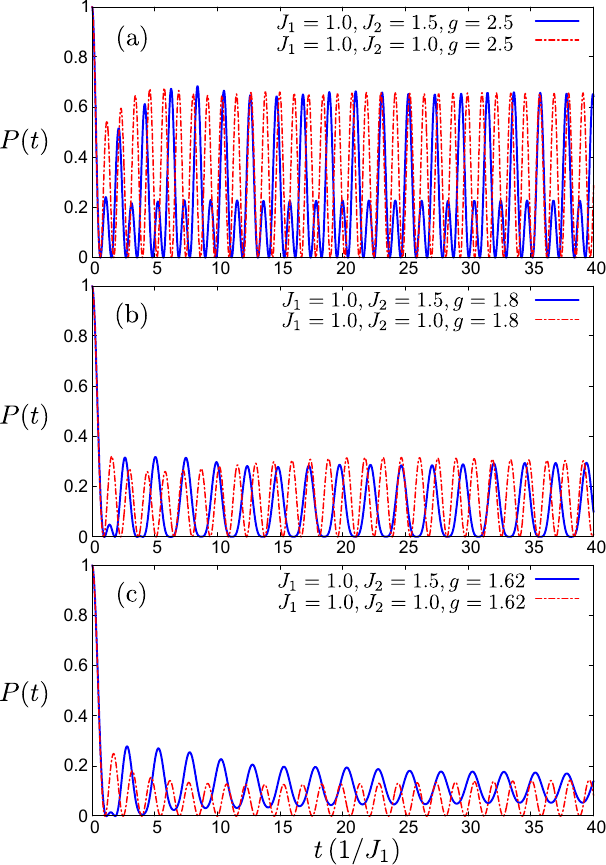}
\hfill
\\
\caption{ Dynamics in Region I when the bath is in the trivial configuration $J_1 < J_2$ and all three bound states are present (blue curves) for three cases.  
We further indicate the corresponding evolution for the uniform, tight-binding chain with 
$J_1 = J_2$ by the red dash-dotted curve in each case (the zero mode is absent for the uniform chain).
}
 \label{fig:dyn.reg.I}
 \end{figure}

For Region I in the case $J_1 < J_2$, the spectrum includes three bound states: the two strong-coupling bound states and the zero-energy bound state, which are the primary influence on the dynamics in this case.
We show several numerical simulations of the dynamics for this case in Fig. \ref{fig:dyn.reg.I}. 

We can obtain an analytic picture for the dynamics working from the expression for the survival amplitude in Eq. (\ref{amp.green}) by dragging the integration contour $C$ down to infinity in the lower half of the complex energy plane as shown in Fig. \ref{fig:contour}(b).  Due to the exponential in the integrand in Eq. (\ref{amp.green}), the contour directly at infinity vanishes, leaving us with the pole contributions $A_0 (t)$ and $A_\pm (t)$ from the three bound states $E_0$ and $E_\pm$ as well as the non-vanishing contributions $\Abr (t) = \sum_{i=1}^4 A_n(t)$ left over from the branch points appearing at each edge of the two continuous energy bands $E_{k_\pm}$ [see Fig. \ref{fig:contour}(b)].  These latter contributions give the branch-point effect, which yields non-Markovian, power-law decay \cite{GNOS19,BCP10,PastCPL06,Khalfin,Sudarshan77,GPSS13}.  
Hence, in Region I the survival amplitude can be written as the contributions
\beq
  A(t) = A_0 (t) + A_+ (t) + A_- (t) + \Abr (t)
  	.
\label{RI.contributions}
\eeq
The pole contributions are easily evaluated from the residue theorem;
the zero mode $z = E_0$ contribution gives
\beq
  A_0 (t) = \frac{J_2^2 - J_1^2}{J_2^2 - J_1^2 + g^2}
  	.
\label{1A.A0}
\eeq
Notice that this quantity would vanish in the limit $J_2 \to J_1$, as it must since the zero mode disappears when the band gap closes.  We similarly calculate the pole contributions from the $z = E_\pm$ bound states, which give
\beq
  A_\pm (t) = \frac{J_1^2 J_2^2 - \left( J_1^2 - g^2 \right)^2}
  					{2 \left( g^2 - J_1^2 \right) \left[ g^2 - \left( J_1^2 - J_2^2 \right) \right]} e^{- i z_\pm t}
	.
\label{1A.Apm}
\eeq
Combining these three pole contributions in Region I for $J_2 > J_1$ we obtain
\begin{widetext}
\beqa
   A_0 (t) + A_+ (t) + A_- (t)	
  	= \frac{\left( J_1^2 J_2^2 - \left( J_1^2 - g^2 \right)^2 \right) \cos z_+ t 
									+ \left( g^2 - J_1^2 \right) \left( J_2^2 - J_1^2 \right)}
  					{\left( g^2 - J_1^2 \right) \left[ g^2 - \left( J_1^2 - J_2^2 \right) \right]}
\label{1A.RI}
\eeqa
\end{widetext}
This contribution clearly gives rise to the oscillations seen in each panel of Fig. \ref{fig:dyn.reg.I} (blue curves). The key point to understanding the oscillations in each case is whether the first or second term in the numerator of Eq. (\ref{1A.RI}) is dominant.  For $g > \sqrt{J_1^2 + J_2^2}$, the maximum value of the first term (hybridized bound states) is larger than the magnitude of the second (zero mode), which results in a simple beat pattern as seen in Fig. \ref{fig:dyn.reg.I}(a)  for $J_1 = 1$, $J_2 = 1.5$ and $g = 2.5$.
However, if we decrease $g$ such that $\sqrt{J_1 (J_1 + J_2)} < g < \sqrt{J_1^2 + J_2^2}$, the second term in Eq. (\ref{1A.RI}) becomes dominant, so that the oscillating term can never completely cancel it out.  Hence, the probability at the emitter never vanishes completely as shown for $g = 1.62$ in Fig. \ref{fig:dyn.reg.I} (c).
Notice that as we decrease the coupling $g$ throughout the panels in Fig. \ref{fig:dyn.reg.I}, the role of the zero-energy bound state becomes more pronounced, which is opposite of the behavior of the strong-coupling bound states.

In each panel of Fig. \ref{fig:dyn.reg.I} (a, b, c) we also show the comparable evolution for the uniform chain with $J_1 = J_2 = 1$ as the red dash-dotted curve.  One can easily observe that the dynamics are qualitatively similar, after one accounts for the absence of the zero mode in the uniform chain.

\begin{figure}
 \includegraphics[width=0.45\textwidth]{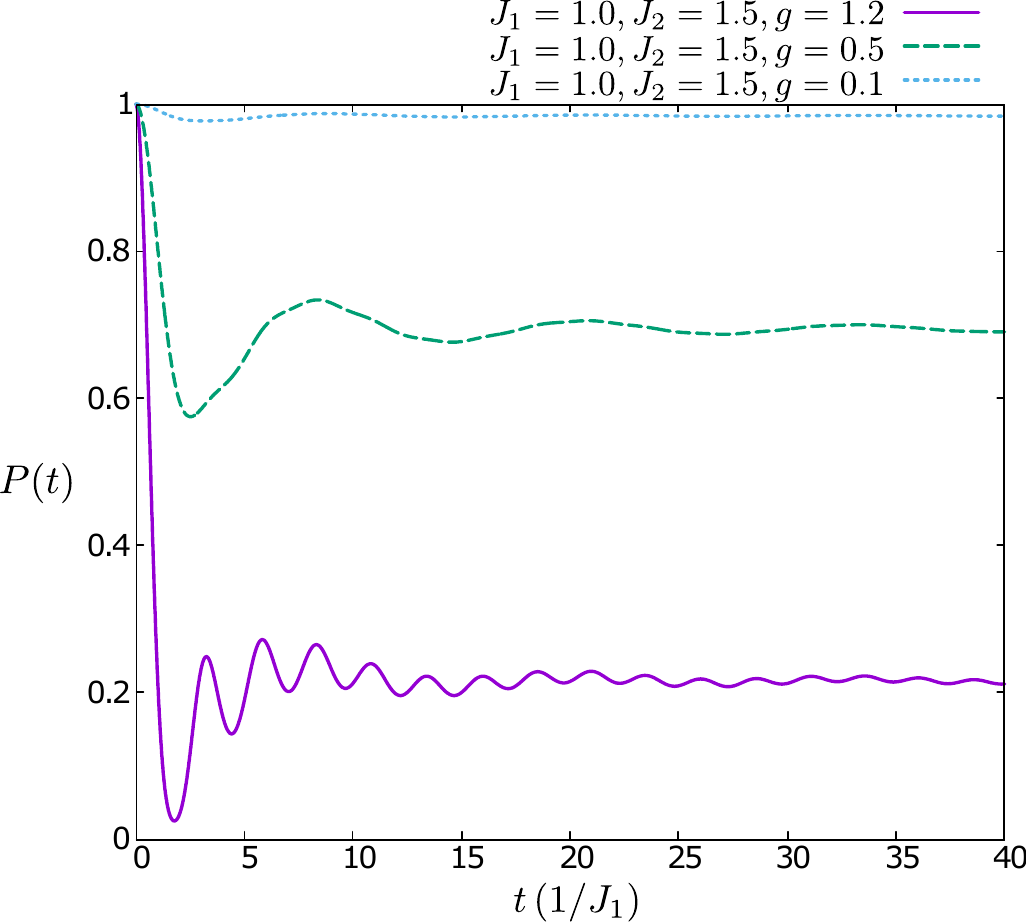}
\hfill
\\
\caption{ Dynamics in Regions II and III when the bath is in the trivial configuration $J_1 < J_2$.  We present the Region II case $g=1.2$ and Region III cases $g = 0.5$ and $g=0.1$.}
 \label{fig:dyn.reg.triv}
 \end{figure}

Moving on, as we continue to decrease the value of $g$ the two outer-gap bound states approach their respective nearby band edges and, at the exact point $g = \gstr$,
 they both touch the band edge before transitioning into delocalized states that reside in the higher Riemann sheet \cite{GPSS13}.  This marks the transition into Region II.  We present the dynamics for a representative case in Region II as the purple curve in Fig. \ref{fig:dyn.reg.triv} for $g = 1.2$.  One might at first be surprised to notice that the oscillations remain in the dynamics even though the two outer bound states have vanished.  The origin of the oscillations in this case, however, are a result of interference between the branch-point contributions from the two outer band edges, similar to Ref. \cite{GNOS19}.  This is worked out for the present case in detail in App. \ref{app:1A.BPE}.

Finally, we move into Region III, which is the weak-coupling regime for $J_1 < J_2$.  Here, the anti-bound states from Region II have transitioned into a resonance/anti-resonance pair.  In many cases, the resonance is associated with exponential decay \cite{PPT91,Sudarshan78,HSNP08,OH17A,Madrid24}; however, in the present scenario in which the gap is fairly wide [recall Fig. \ref{fig:1A.spec.triv}(a)], the resonance plays little role in the dynamics.  Instead, the survival probability exhibits fractional decay associated with the zero-energy mode following a brief period of decay resulting from the branch-point contributions.  This is shown for the cases $g = 0.5$ (green dashed curve) and $g = 0.1$ (blue dotted curve) in Fig.  \ref{fig:dyn.reg.triv}.  Notice that the asymptotic probability of decay decreases steadily as the value of $g$ is decreased, and the system becomes more stable.  This demonstrates the influence of the first type of weak-coupling bound state in the present system.  However, notice the dynamics are arguably somewhat trivial, since the particle is largely trapped in the quantum emitter.

Finally, we note that the frequency of the oscillations appearing during the early, decaying part of the evolution decreases as we increase $g$ in Fig.  \ref{fig:dyn.reg.triv}.  This is because the inner band-edge contributions to the branch-point effect become more important than the outer band-edge contributions as we move deeper into Region III.


\subsection{$J_1 > J_2$ case: Dynamics in Region V}\label{sec:1A.dynamics.topological}

Here we study the dynamics in the case $J_1 > J_2$, for which the bare reservoir (now in the topological phase) exhibits the surface state given in Eq. (\ref{semi.edge}).  The spectrum for the full model in this case was shown in Fig. \ref{fig:1A.spec.top}.  We briefly comment on Regions I-IV, but our primary focus will be Region V.

The dynamics for Regions I, II and III for $J_1 > J_2$ are qualitatively similar to what we have previously shown for the corresponding regions from the $J_1 < J_2$ case, except that the zero mode bound state is now absent.  Further, as long as the gap $J_1 - J_2$ is kept relatively wide, the dynamics in Region IV are not much different than Region III (the narrowband case may be presented in future work).  

\begin{figure}
 \includegraphics[width=0.45\textwidth]{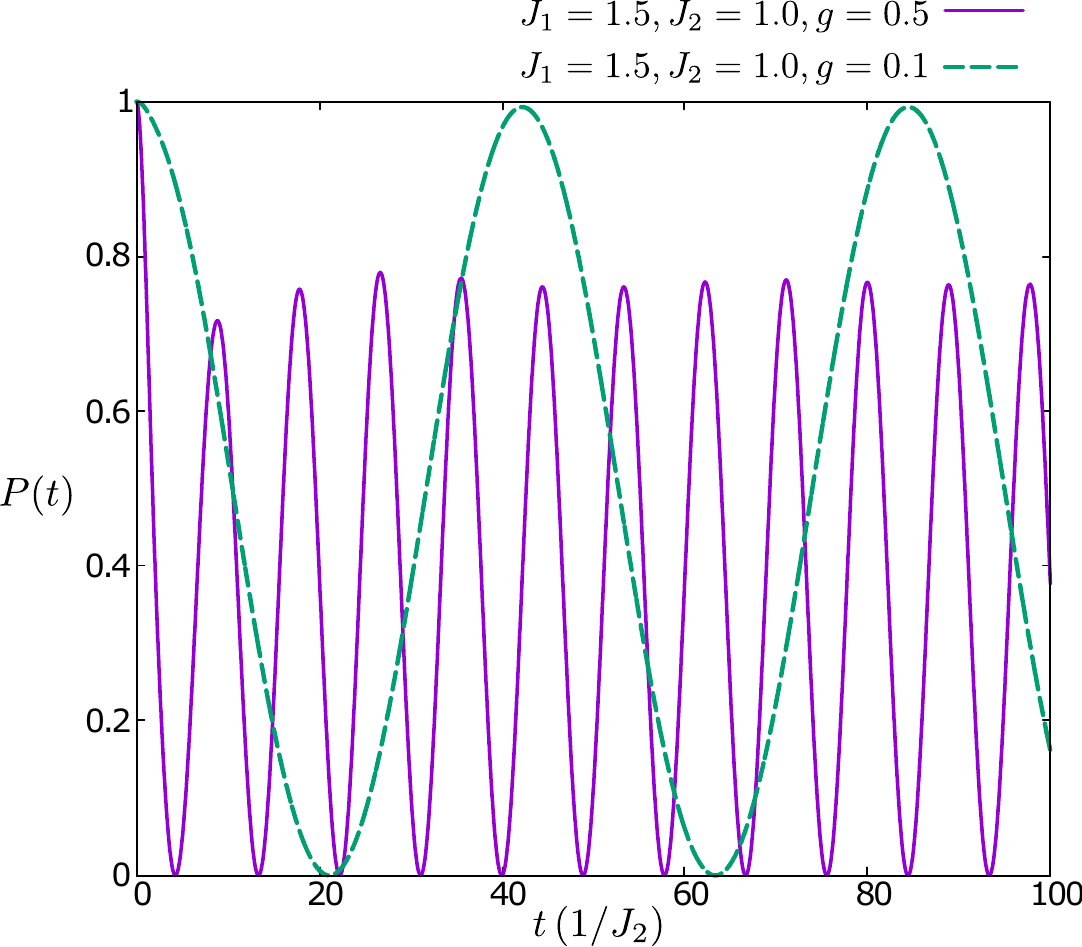}
\hfill
\\
\caption{ Rabi oscillations in Regions V when the bath is in the topological configuration $J_1 > J_2$ with $J_1=1.5$ and $J_2 =1.0$.  We present two cases, $g = 0.5$ (full purple curve) and $g=0.1$ (green dash-dotted curve).}
 \label{fig:dyn.reg.V}
 \end{figure}

For Region V, the distinguishing dynamical feature is of course the two weak-coupling bound states, which give rise to a Rabi oscillation as shown for the cases $g=0.5$ and $g = 0.1$ in Fig. \ref{fig:dyn.reg.V} (with $J_1 = 1.5, J_2 = 1.0$).  Because the Region V bound states are located within the gap, the frequency of the oscillations is much slower compared to Region I.  In particular, assuming $g \ll J_1$, the oscillation frequency is on the order $\sim 2 g \sqrt{1 - (J_2/J_1)^2}$.

So far, we have only considered the dynamics of the quantum emitter itself.  Mainly this is because we would expect the particle to either be trapped inside the emitter or to be dissipated into the reservoir in a fairly uniform way under the previous scenarios.
However, the weak-coupling bound states give rise to a simple, coherent evolution, in which the the future location of the particle on one of the two sublattices within the chain is easily predicted and the particle is recoverable at the emitter.

\begin{figure}
 \includegraphics[width=0.45\textwidth]{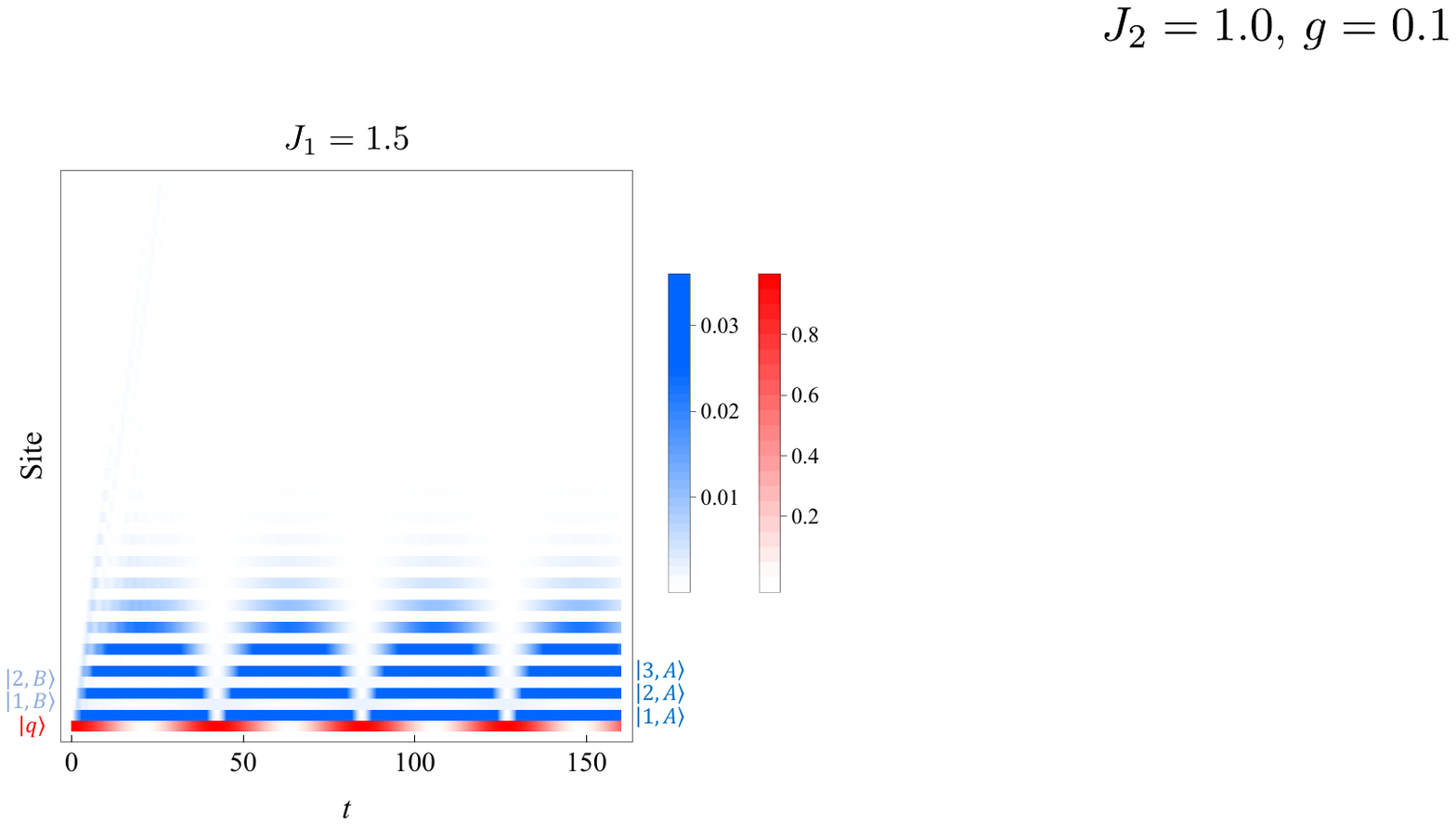}
\hfill
\\
\caption{Evolution of the probability distribution $\left| \bra x | e^{-iHt} | q \ket \right|^2$ with site $\bra x | = \bra q |$, $\bra n, A |$ or $\bra n, B |$
for the wide-band case with $J_2 = 1.5$, $J_1 = 1.0$, and $g=0.1$.
}
 \label{fig:osc15}
 \end{figure}

In particular, the symmetric and anti-symmetric combination states $| \psi_{s,a} \ket$ we introduced in Eq. (\ref{psi.sa}) are useful for gaining insight into the evolution in this situation.  Expanding these states for small $g$ under the assumption $g^2 \ll J_1^2 - J_2^2$, we find
\beq
  | \psi_s \ket 
  	\approx | q \ket - \frac{g J_2}{J_1^2} \sum_{n=1}^\infty \left( - \frac{J_2}{J_1} \right)^{n-1} | n, B \ket 
		+ \mathcal{O}(g^2)
\label{psi.s.g}
\eeq
and
\beq
  | \psi_a \ket 
  	\approx \frac{\sqrt{J_1^2 - J_2^2}}{J_1} \sum_{n=1}^\infty \left( - \frac{J_2}{J_1} \right)^{n-1} | n, A \ket
		+ \mathcal{O}(g^2)
	.
\eeq
Notice that the lowest-order expression in $| \psi_a \ket$ is exactly the edge state of the decoupled SSH chain from Eq. (\ref{semi.edge}).  Hence, to lowest-order $| \psi_a \ket $ behaves similar to the edge state of the bare reservoir while $| \psi_s \ket$ behaves similar to the quantum emitter itself.  As a result, we can use Eq. (\ref{psi.sa.time}) to approximate the quantum emitter evolution as
\begin{widetext}
\beq
  e^{-i H t}  | q \ket 
  	\approx  \cos \left[ \frac{g \sqrt{J_1^2 - J_2^2}}{J_1} t \right] \ | q \ket  
		- i \sin \left[ \frac{g \sqrt{J_1^2 - J_2^2}}{J_1} t \right] \ | \phi_{e} \ket + | \psi_\textrm{FM} (t) \ket 
\label{weak.g.evol}
\eeq
\end{widetext}
on the timescale $g^2 t \lesssim 1$, where $| \psi_\textrm{FM} (t) \ket$ gives a free-motion like evolution, to which we will return shortly.
Notice that without the term $ | \psi_\textrm{FM} (t) \ket $, Eq. (\ref{weak.g.evol}) represents a particularly simple evolution in which the probability to find the particle oscillates from the initialized $|q\ket$ site into the edge state of the bare SSH chain and then back again before repeating.  
Indeed, the evolution shown in Fig. \ref{fig:osc15} for the case $J_1= 1.5, J_2 = 1.0$ and $g=0.1$ conforms to this description quite well.  However, notice that the particle only penetrates to about the 5th cell of the chain.

\begin{figure}
 \includegraphics[width=0.45\textwidth]{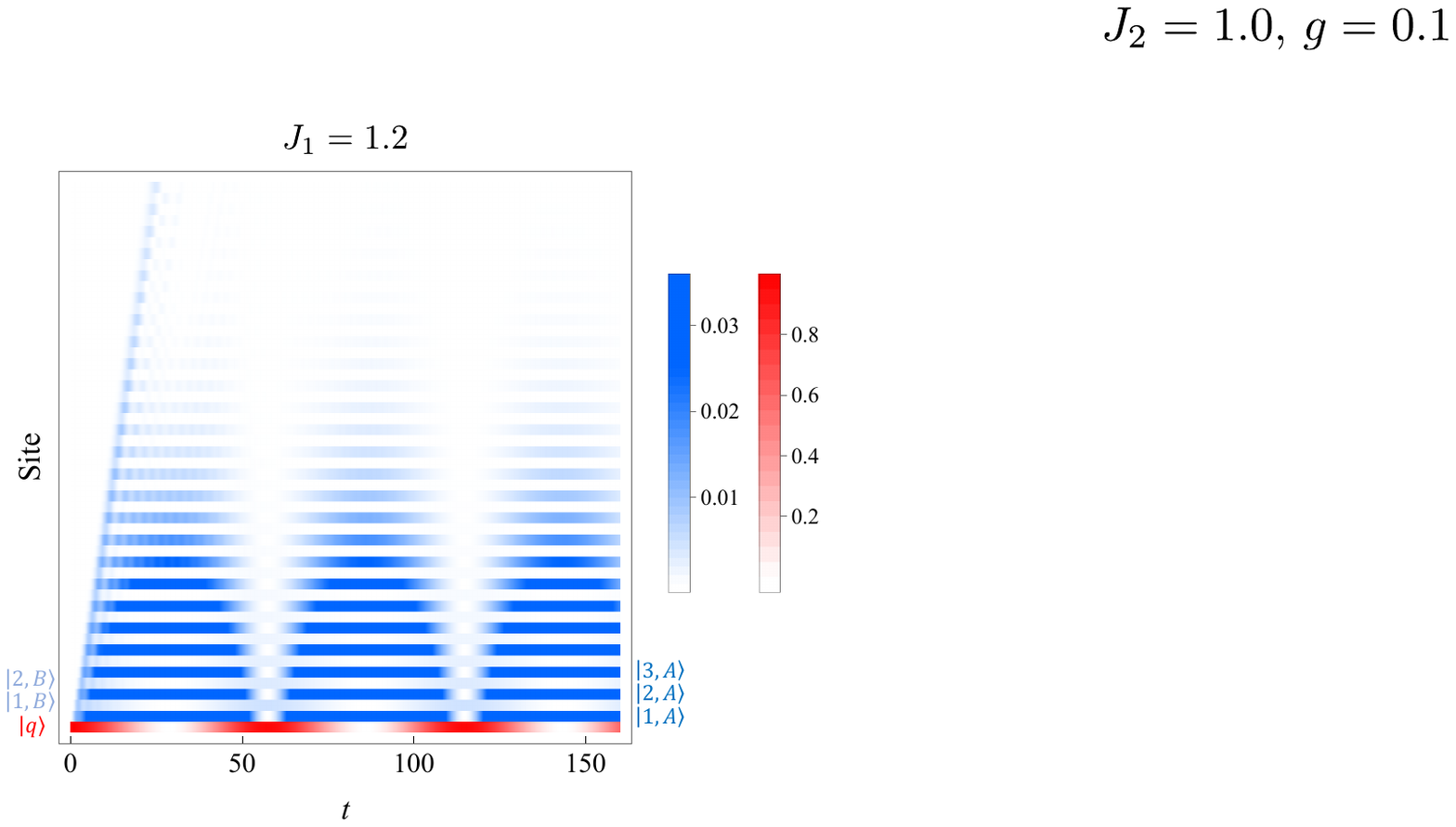}
\hfill
\\
\caption{Evolution of the probability distribution $\left| \bra x | e^{-iHt} | q \ket \right|^2$ with site $\bra x | = \bra q |$, $\bra n, A |$ or $\bra n, B |$
 for the case $J_2 = 1.2$, $J_1 = 1.0$, and $g=0.1$.
}
 \label{fig:osc12}
 \end{figure}

The form of the equation for the edge state in (\ref{semi.edge}) makes clear that to achieve greater lattice penetration, one needs to narrow the gap given by $J_1 - J_2$.  However, there is a tradeoff: the term
$| \psi_\textrm{FM} (t) \ket$, which gives a free-motion like evolution that is the primary mechanism of population transfer to the $B$ sublattice, 
acquires greater influence as one closes the gap.   A careful analysis reveals that the condition for neglecting $| \psi_\textrm{FM} (t) \ket$ is roughly given by $g < J_1 - J_2$, which provides guidance on how much we can narrow the gap relative to the coupling strength [see App. \ref{app:1A.FM} for a detailed analysis]. The effect of narrowing the gap  is demonstrated for the case $J_1= 1.2, J_2 = 1.0$ and $g=0.1$ in Fig. \ref{fig:osc12}, in which case the particle can reach about twice as far into the lattice; however, the influence of the $| \psi_\textrm{FM} (t) \ket$ term can be seen as it somewhat distorts the first oscillation. 

\begin{figure}
 \includegraphics[width=0.45\textwidth]{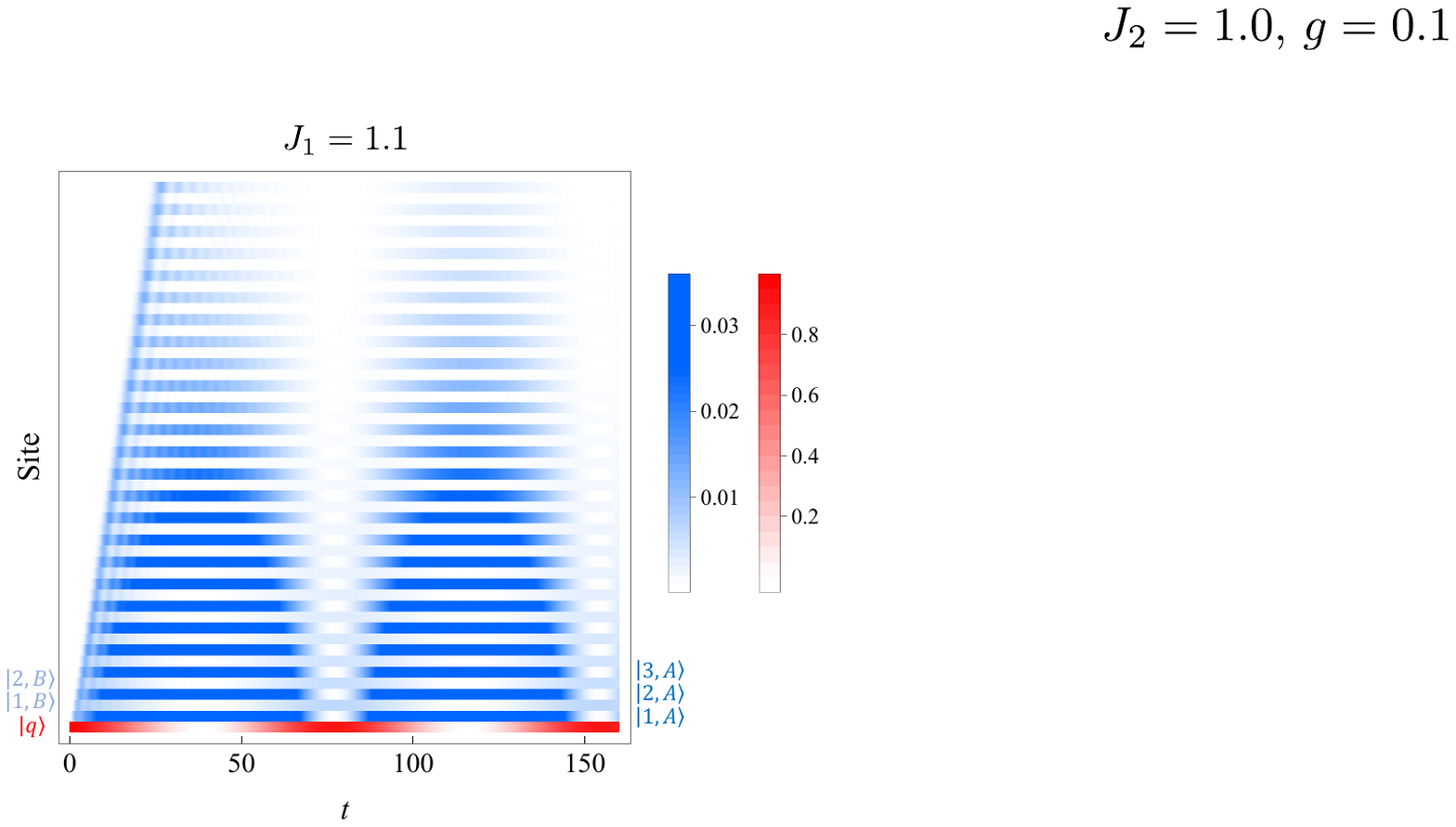}
\hfill
\\
\caption{Evolution of the probability distribution $\left| \bra x | e^{-iHt} | q \ket \right|^2$ with site $\bra x | = \bra q |$, $\bra n, A |$ or $\bra n, B |$
 for the narrow-band case $J_2 = 1.1$, $J_1 = 1.0$, and $g=0.1$.
}
 \label{fig:osc11}
 \end{figure}

Next, in Fig.  \ref{fig:osc11} we show the case  $J_1= 1.1, J_2 = 1.0$ and $g=0.1$, for which the exact condition $g=J_1 - J_2$ is satisfied.  In this case, we see that the free motion-like component of the evolution is much more prominent, essentially extending to infinity, although the oscillations that are our primary interest are still entirely coherent.  As detailed in App.  \ref{app:1A.FM}, the free motion-like evolution can be described as a wave packet with group velocity $v_g = J_2$ that gradually overpowers the edge-state oscillations as one closes the gap in the limit $J_1 \to J_2$, which explains how the dynamics of the uniform chain is recovered in this limit.

\section{Conclusion}\label{sec:conclusion}

In this work, we have revealed the existence of two types of weak-coupling bound states in a model consisting of a quantum emitter coupled to a semi-infinite extension of the SSH model.  The first of these was essentially just a modified version of the edge state from the original SSH model that now is localized around the quantum emitter site.  Meanwhile the second type of weak-coupling bound state appeared instead as a pair of states hybridized between the quantum emitter and the edge state from the bare SSH reservoir.  These latter states are qualitatively similar to the bound states studied numerically in the 2-D model in Ref. \cite{Zhang22}, 
but here we have provided a detailed profile of their analytic properties.
Further, we have shown that these states can be used as a control mechanism; specifically, as a way to populate the edge state of the bare SSH chain.  Alternatively, we could view this process as a means to transfer the particle from the emitter site to sites specifically on the $A$ sublattice of the bare chain, but we emphasize that the particle can later be recovered at the emitter.

We mention that we have performed preliminary investigations of a model in which the semi-infinite SSH reservoir from the present work is replaced with a 2-D extension of the SSH model \cite{Asboth2016}, in which we have again discovered the presence of the hybridized weak-coupling bound states (our 2-D model with the quantum emitter appears in the previous work Ref. \cite{Josh23}).  Hence, we feel confident in claiming these states should be a somewhat general feature of models that include a quantum emitter coupled with the surface state of a topological structured reservoir.  Such an arrangement might provide a mechanism for controlling particle transport in topological insulators.

Let us briefly return to the question of how to interpret the winding number(s) introduced in the models considered in this paper.  We recall that in the original concept of bulk-boundary correspondence, the non-zero winding number derived from the bulk eigenstates in the model with periodic boundary conditions is used to predict the existence of edge states in essentially the same model, but after imposing open boundary conditions.  In the present case, our semi-infinite model constitutes a kind of hybrid geometry, in which we have continuum (``bulk'') eigenstates from the infinite side of the chain, while we maintain an effective edge on the finite side (although the boundary condition itself is open on either end).  Hence, in this case we found that the non-zero winding number of the continuum eigenstates more simply predicts the existence of the (edge-like) bound states in the model with the {\it same} boundary conditions.  Hence, we thought of this as a kind of {\it in-house bulk-boundary correspondence}.

Finally, we note that in this work we have only considered the case in which the emitter couples to the edge state, directly at the endpoint site of the semi-infinite SSH chain.  In particular, we have not considered the situation in which the emitter couples with one of the $B$ sites along the chain, for which the amplitude of the edge state vanishes [recall Eq. (\ref{semi.edge})].  In this case, localized states would form in which the emitter hybridizes with a small subset of sites on the finite side of the SSH chain, but has vanishing amplitude everywhere else.  These can be considered examples of compact localized states \cite{Sutherland,Feinberg96,Flach17,Kempkes23} or, more generally, the vacancy-like dressed states from Ref.  \cite{LCC21}.
In contrast with the weak-coupling bound states studied in this work, the energy of these vacancy-like dressed states are entirely insensitive to the emitter coupling strength (although their wave function and other physical properties are not).  These states will be the subject of future work.


\section*{Acknowledgments}
This work was supported by the Japan Society for the Promotion of Science under KAKENHI Grant Nos. JP18K03466 and JP22K03473.  We also acknowledge insightful discussions with J. Feinberg, V. Jagadish, K. Kanki, F. Roccati, and S. Shirai.


\appendix

\section{Calculation of the wave function for the bound states}\label{app:bs}

In this appendix, we obtain the wave function for the bound states in Regions I and V, which we write as $| \psiBpm \ket$.
Starting from the eigenvalue equation for the bound states 
$H | \psiBpm \ket = z_\pm| \psiBpm \ket$
we can write a relationship between the quantum emitter component of the eigenstate $\bra q | \psiBpm \ket$
and the components in the first cell of the SSH chain as
\beq
	 \left[  \begin{array}{c}
		\bra 1, A|\psiBpm \ket   	\\
		\bra 1, B|\psiBpm \ket 
			\end{array} \right] 
			= \left[  \begin{array}{c}
		z_\pm/g  	\\
		\frac{z_\pm^2/g -g}{J_2}
			\end{array} \right] 
				\bra q | \psiBpm \ket 
	.
\label{app2_q}
\eeq
Similarly, we find a recursive relation between the amplitudes of cell $n+1$ in terms of those for the cell $n$ as
\beq
\left[  \begin{array}{c}
		\bra n+1, A|\psiBpm \ket  	\\
		\bra n+1, B|\psiBpm\ket  
			\end{array} \right]
=
M_\pm
	\left[  \begin{array}{c}
		\bra n, A|\psiBpm\ket   	\\
		\bra n, B|\psiBpm\ket 
			\end{array} \right]
\label{app2_2}
\eeq
for $n\ge 1$,  where 
\beq
M_\pm = \left[  \begin{array}{cc}
		-J_2/J_1				& z_\pm/J_1			\\
		-z_\pm/J_1	& \frac{z_\pm^2-J_1^2}{J_1 J_2}	
		 	\end{array} \right]
	.
\eeq
Thus, we find
\beq
\left[  \begin{array}{c}
		\bra n+1, A|\psiBpm\ket  	\\
		\bra n+1, B|\psiBpm\ket  
			\end{array} \right]
=
M_\pm^n
	\left[  \begin{array}{c}
		\bra 1, A|\psiBpm\ket   	\\
		\bra 1, B|\psiBpm\ket 
			\end{array} \right].
\label{app2_3}
\eeq
To evaluate the $n^{\rm th}$ power of the matrix $M_\pm$, we will use its right and left eigenvectors,  $\xi_{j\pm}$ and ${\tilde\xi}_{j\pm}^T$ (with $j=1,2$),  satisfying 
\beq
\begin{array}{c}
		M_\pm \xi_{j\pm} = m_{j\pm} \xi_{j\pm}\\
		 {\tilde\xi}_{j\pm}^T M_\pm = {\tilde\xi}_{j\pm}^T m_{j\pm}
		 			\end{array}
\eeq
where $m_{j\pm}$ are the eigenvalues. Using the expression for the bound-state eigenvalues $z_\pm = E_\pm$ from Eq. (\ref{E.pm}),
we find 
\beq
\begin{array}{ccc}
		m_{1\pm}=-J_2/{\tilde J}_1, & \xi_{1\pm} = {\cal N} \left[\begin{array}{c} 1 \\ \mp r \end{array}\right], & {\tilde\xi}_{1\pm}^T= {\cal N} \left[\begin{array}{cc}1 & \pm r \end{array}\right]\\ \\
		 m_{2\pm} = -{\tilde J}_1/J_2,  & \xi_{2\pm} = {\cal N} \left[\begin{array}{c} \mp r \\ 1 \end{array}\right], & {\tilde\xi}_{2\pm}^T= {\cal N} \left[\begin{array}{cc}\pm r & 1 \end{array}\right]
			\end{array}
\eeq
where
\beq
		 {\tilde J}_1 = J_1-g^2/J_1,
\eeq
\beq
		 r = \frac{gJ_2}{J_1^2-g^2}\sqrt{\frac{J_1^2-g^2}{J_1^2-J_2^2-g^2}},
\eeq
or
\beq
		  r= \pm \frac{g J_2}{\sqrt{(J_1^2 - J_2^2 - g^2)(J_1^2-g^2)}} \quad {\rm for}\, J_1\gtrless g
\eeq
and
\beq
		 {\cal N} =  \frac{1}{\sqrt{1-r^2}}
	.
\eeq
Since the eigenvectors of $M_\pm$ form a complete orthonormal set, 
\beq
	{\tilde\xi}_{j'\pm}^T	\xi_{j\pm} = \delta_{j'j}
	\label{app2_5} 
\eeq
\beq
	\sum_j	\xi_{j\pm} {\tilde\xi}_{j\pm}^T= I_2	 
\eeq
where $I_2$ is the $2\times 2$ identity matrix, we can evaluate $M_\pm^n$ as
\beq
	 M_\pm^n = \sum_j \xi_{j\pm} m_j^n {\tilde\xi}_{j\pm}^T
	 \label{app2_4}
\eeq
where $m_{j+} = m_{j-}=m_j$. Finally, from the relation for the quantum emitter amplitude (\ref{app2_q}) we find
\beq
	 \left[  \begin{array}{c}
		\bra 1, A|\psiBpm\ket   	\\
		\bra 1, B|\psiBpm\ket 
			\end{array} \right] 
			= \bra q |\psiBpm \ket \frac{z_\pm}{g\cal{N}}\xi_{1\pm},
\eeq
which, with Eqs. (\ref{app2_3}), (\ref{app2_4}) and the orthonormality relation (\ref{app2_5}) give
\beq
\left[  \begin{array}{c}
		\bra n+1, A|\psiBpm\ket  	\\
		\bra n+1, B|\psiBpm\ket  
			\end{array} \right] = \bra q |\psiBpm\ket \frac{z_\pm}{g} \left(-\frac{J_2}{{\tilde J}_1}\right)^n\left[\begin{array}{c} 1 \\ \mp r \end{array}\right] .
			\label{app2_7}
\eeq
This gives the $A$ and $B$ components of the bound eigenstates along the lattice as reported in the main text, Eq. (\ref{psi.BS}). From the normalization condition of these states we find
\beq
\bra q |\psiBpm\ket = \left[1 + \left( \frac{z_\pm}{g}\right)^2 \left( 1+ r^2\right) \frac{1}{1-\left(J_2/{\tilde J}_1\right)^2}\right]^{-1/2}
	.
\label{app2_6}
\eeq


\section{Evaluation of the Green's function by the method of continued fractions}\label{app:1A.green}

In this appendix, we evaluate the Green's function for quantum emitter $| q \ket$ attached to the semi-infinite SSH chain, for which we introduce the shorthand notation
\beq
  \alpha_q \equiv  \bra q | \frac{1}{z - H} | q \ket
  	.
\label{alpha.q}
\eeq
For ease of reference, we also write the full Hamiltonian from Eq. (\ref{1A.ham.2q}) in the explicit single-particle representation
\beqa
  H
	& = & g  | q \ket \bra 1, a |
  					+ g | 1, a \ket \bra q |
					\nonumber		\\
	& & +	\sum_{n=1}^\infty \left[ J_2 \left( | n, A \ket \bra n, B | + | n, B \ket \bra n, A | \right) \right.
  			\label{1A.ham}	\\
  	& &	\left. + J_1 \left(  | n, B \ket \bra n+1, A | + | n+1, A \ket \bra n, B | \right)	 \right]
				\nonumber
	.
\eeqa

To evaluate the Green's function in Eq. (\ref{alpha.q}), we apply the method of continued fractions \cite{DBP08}
(some other, possibly related methods that can be applied in semi-infinite systems can be found in Refs. \cite{GomezL17,GomezL22,Znojil25}).  
As a starting point, let us act on $| q \ket$ from the left with $z-H$, which gives
  $\left( z - H \right) | q \ket = z | q \ket - g | 1, a \ket$
  	.
Closing this with $ \bra q | (z-H)^{-1}$ we obtain
\beq
  1 = z \; \alpha_q - g \; \alpha_{1,A}
  	,
\label{e.e.2}
\eeq
in which we have introduced the further shorthand notation
\beq
  \alpha_{n,X} \equiv  \bra q | \frac{1}{z - H} | n, X \ket
\eeq
where $n$ is the $n^{\rm th}$ unit cell and $X = A$ or $B$.
Factoring out $\alpha_q$ on the RHS of (\ref{e.e.2}) and re-arranging allows us to write this equation in the recurrent form
\beq
  \alpha_q = \frac{1}{ z - g \frac{\alpha_{1,A}}{\alpha_q} }
  	.
\label{e.e.r}
\eeq

Next, we perform a similar analysis for the $| 1, A \ket$ state.  We first evaluate
 $\left( z - H \right) | 1, A \ket = z |1, A \ket - g | q \ket - J_2 | 1, B \ket$.
After closing this with $ \bra q | (z-H)^{-1}$ 
we can obtain
\beq
  g \frac{\alpha_q}{\alpha_{1,A}} = z - J_2 \frac{\alpha_{1,B}}{\alpha_{1,A}}
  	.
\label{1a.4}
\eeq
This can be written in the recurrent form
\beq
  \alpha_{1,A} = \frac{g \alpha_{q} }{ z - J_2 \frac{\alpha_{1,B}}{\alpha_{1,A}} }
  	.
\label{1a.r}
\eeq

A similar evaluation can be performed for an arbitrary site $n$ on the $B$ sublattice to obtain
\beq
  \alpha_{n,B} = \frac{J_2 \alpha_{n,A} }{ z - J_1 \frac{\alpha_{n+1,A}}{\alpha_{n,B}} } ,
  	\ \ \ \ \ \ \ \ \ \ \ \ \ \  n \ge 1
  	,
\label{nb.r}
\eeq
as well as for a site $n > 1$ on the $A$ sublattice, which gives
\beq
  \alpha_{n,A} = \frac{J_1 \alpha_{n-1,B} }{ z - J_2 \frac{\alpha_{n,B}}{\alpha_{n,A}} } ,
  	\ \ \ \ \ \ \ \ \ \ \ \ \ \  n > 1
  	.
\label{na.r}
\eeq
We can now rewrite Eq. (\ref{e.e.r}) for the desired Green's function as
\beq
  \alpha_q
	= \frac{1}{ z - \frac{g^2}{  z - J_2 \frac{\alpha_{1,B}}{\alpha_{1,A}} } }
	= \frac{1}{ z - \frac{g^2}{  z -  \frac{J_2^2}{ z - J_1 \frac{\alpha_{2,A}}{\alpha_{1,B}} } } }
  	,
\label{e.e.r.2}
\eeq
where we have inserted Eq. (\ref{1a.r}) in the first step, and then inserted (\ref{nb.r}) in the second step.
From this point, one alternately applies Eq. (\ref{na.r}) then (\ref{nb.r}) successively, to obtain the Green's function as
\beq
  \alpha_q = \frac{1}{ z - \frac{g^2}{  z - \Xi_\textrm{SSH} (z) } }
\label{e.e.ssh}
\eeq
in which
\beq
  \Xi_\textrm{SSH}
  	= \frac{J_2^2}{z - \frac{J_1^2}{z - \frac{J_2^2}{z - \frac{J_1^2}{z - \dots}} }}
	= \frac{J_2^2}{z - \frac{J_1^2}{z - \Xi_\textrm{SSH} }}
\label{ssh.self.r}
\eeq
is the self-energy function of the SSH reservoir.  This latter equation can be solved for
\beq
  \Xi_\textrm{SSH}
  	= \frac{z^2 - (J_1^2 - J_2^2) \pm \sqrt{z^4 - 2z^2 (J_1^2 + J_2^2) + (J_1^2 - J_2^2)^2 } }{2 z}
	.
\label{ssh.self}
\eeq
Applying this finally in Eq. (\ref{e.e.ssh}) gives the expressions Eq. (\ref{1A.q.q}) and (\ref{1A.self}) from the main text.


\section{Evaluation of the branch-point effect in Region II}\label{app:1A.BPE}

In this appendix, we outline the evaluation of the non-Markovian dynamics from the branch-point effect presented in Region II, Sec. \ref{sec:1A.dynamics.trivial} of the main text.  First, based on the contour deformation shown in Fig. \ref{fig:contour} , we can break the integration for the branch-point contributions into four parts, each associated with one of the branch points at the band edges.  This takes the form
\beq
  \Abr (t) = A_1 (t) + A_2 (t) + A_3 (t) + A_4 (t)
  	,
\label{br.conts}
\eeq
in which $A_{1,4} (t)$ are the contributions from the outer band edges, while $A_{2,3} (t)$ are those from the inner band edges.

In the following, we evaluate the contribution $A_4 (t)$ associated with the uppermost band edge as an example.  The other contributions work out similarly.  
First, we apply the expression for the Green's function from Eqs. (\ref{1A.q.q}) and (\ref{1A.self}) in the survival amplitude Eq. (\ref{amp.green}).  Then the contour contribution associated with $A_4 (t)$ in Fig, \ref{fig:contour}(b) can be written in terms of the contour going out to infinity in the second Riemann sheet subtracted from that coming in from infinity on the first sheet, which takes the form
\begin{widetext}
\beqa
  A_4 (t) 
  &	= & \frac{1}{4 \pi i } \int_{J_1 + J_2 - i \infty}^{J_1 + J_2} dz e^{-izt} \
		\frac{(2 J_1^2 - g^2) z^2 -g^2 J_1^2 + g^2 J_2^2 
  						- g^2  \sqrt{z^4 - 2z^2 (J_1^2 + J_2^2) + (J_1^2 - J_2^2)^2 } }
				{z \left[  \left( J_1^2 - g^2 \right) z^2 + g^2 \left( J_2^2 - J_1^2 + g^2 \right)  \right]}  
				\nonumber	\\
  & &	+ \frac{1}{4 \pi i } \int_{J_1 + J_2}^{J_1 + J_2 - i \infty} dz e^{-izt} \ 
		\frac{(2 J_1^2 - g^2) z^2 -g^2 J_1^2 + g^2 J_2^2 
  						+ g^2  \sqrt{z^4 - 2z^2 (J_1^2 + J_2^2) + (J_1^2 - J_2^2)^2 } }
				{z \left[  \left( J_1^2 - g^2 \right) z^2 + g^2 \left( J_2^2 - J_1^2 + g^2 \right)  \right]}
\label{A4.1}					\\
 &	= &  \frac{1}{2 \pi i } \int_{J_1 + J_2}^{J_1 + J_2 - i \infty} dz e^{-izt} \ 
		\frac{ \sqrt{ (z - J_1 - J_2) (z - J_1 + J_2)(z + J_1 - J_2)(z + J_1 + J_2) }}
				{z \left[  \left( J_1^2 - g^2 \right) z^2 + g^2 \left( J_2^2 - J_1^2 + g^2 \right)  \right]}
\label{A4.2}
\eeqa
\end{widetext}
From here, we make the change of integration variables from $z$ to $y$ as implicitly defined by
\beq
  z = J_1 + J_2 - i y
\eeq
in which $y \in \left[ 0 , \infty \right]$ and $dz = -i dy$.  Applying the transformation we can obtain
\begin{widetext}
\beqa
   A_4 (t) 
  	\approx  - \frac{g^2}{2 \pi } e^{-i (J_1 + J_2) t} \int_{0}^{\infty} dy \; e^{-yt} 
		\frac{ \sqrt{8 J_1 J_2 (J_1 + J_2) (-iy)} }
				{ F (y) }
\label{A4.3}
\eeqa
in which the denominator function $F (y)$ takes the form
\beqa
  F (y) & = & (J_1 + J_2) \left[ g^2 - J_1 (J_1 + J_2) \right]^2
					\ - i \left[ g^4 + 3 J_1^2 (J_1 + J_2)^2 - 2g^2 (J_1 + J_2) (2 J_1 + J_2) \right] y
				\nonumber	\\
	& &	+ 3 (g^2 - J_1^2) (J_1 + J_2) y^2  \ - i ( g^2 - J_1^2 ) y^3
	.
\label{F4.y}
\eeqa
Notice in Eq. (\ref{A4.3}) we have dropped terms of order $y^2$ or higher under the radical that are associated with band edges other than the upper one, as the integral is largely insensitive to those details.
We next introduce the change of integration variable $s = yt$, which gives, after slight rearrangement
\beq
   A_4 (t) 
  	\approx - \frac{ g^2 \sqrt{- 8 i J_1 J_2 }}{2 \pi \sqrt{J_1 + J_2} \ t^{3/2}} e^{-i (J_1 + J_2) t} 
			\int_{0}^{\infty} ds \; e^{-s} \frac{ \sqrt{s} }
				{ \left[ g^2 - J_1 (J_1 + J_2)) \right]^2 - i X \frac{s}{t} + \mathcal{O}(s^2/t^2)}
\label{A4.4}	
\eeq
in which
\beq
  X = \frac{g^4 + 3 J_1^2 (J_1 + J_2)^2 - 2g^2 (J_1 + J_2) (2 J_1 + J_2) }{J_1 + J_2}
  	.
\eeq
The denominator in the integrand in Eq. (\ref{A4.4}) is the key point for resolving the non-Markovian dynamics associated with the branch-point effect.  In the case that $t$ is very large, the first term in the denominator tends to be dominant in the integration, with most of the contribution to the integral coming in the range $s \in [0, 1]$ as a result of the overall exponential factor.  Under the specific condition
\beq
  t \gg \frac{X}{\left[ g^2 - J_1 (J_1 + J_2) \right]^2}
		\equiv T_{\Delta_1}
\label{FZ.outer}
\eeq
the first term in the denominator of Eq. (\ref{A4.4}) is dominant.  
Approximating then by throwing out all of the other terms in the denominator, we obtain
\beq
  A_4(t) \approx -  \frac{g^2}{  \left[ g^2 - J_1 (J_1 + J_2) \right]^2 t^{3/2}}  
  				\sqrt{\frac{ - i J_1 J_2}{2 \pi (J_1 + J_2)}} e^{ - i (J_1+J_2) t}
  	.
\label{A4.FZ}
\eeq
\end{widetext}

Finally, combining this with a similar result for the $A_1 (t)$ contribution coming from the lower (outer) band edge, we find
\beq
  \Abr (t)
	\approx - \frac{2 g^2  \alpha_\textrm{FZ,1}}{ \sqrt{\pi} \ t^{3/2}}  
  				 \cos \left( (J_1+J_2) t + \pi/4 \right)
\label{1A.FZ.outer}
\eeq
where the coefficient $\alpha_\textrm{FZ,1}$ is given by
\beq
  \alpha_\textrm{FZ,1}
  	= \frac{1}{ \left[ g^2 - J_1 (J_1 + J_2) \right]^2} \sqrt{\frac{ J_1 J_2}{2 (J_1 + J_2)}}
	.
\eeq
This gives the $1/t^3$ decay in the survival probability $P(t)$ on long times.  
This is the equivalent of the far zone dynamics from Refs. \cite{GPSS13,GNOS19}.
In the case that the timescale $T_{\Delta_1}$ is not much different from (or even less than) the early Zeno timescale $T_\textrm{Z}$, then the early time parabolic decay transitions almost directly into this asymptotic $1/t^3$ decay.  
For example, in Region III this is a quite safe assumption.



However, if instead we consider the parameter space near the transition between Regions II and I, we find that the timescale $T_{\Delta_1}$ becomes well-separated from $T_\textrm{Z}$, such that a new time domain appears in between the two timescales \cite{GPSS13}.  Specifically, if we are on the Region II (Region I) side of the transition, then the anti-bound state (bound state) lying near the outermost band implies the gap $\Delta_1$ between them is small and $T_{\Delta_1}$ is large such that $T_{\Delta_1} \gg T_\textrm{Z}$.  
The calculation is very similar to that above for the asymptotic regime, but instead of the first term in the denominator of Eq. (\ref{A4.4}), we must instead consider the second term in the denominator (linear in $s/t$).  This calculation can be carried out in a similar manner to that above to obtain
\begin{widetext}
\beq
  A_4(t) \approx \frac{ g^2 \sqrt{- 2 i J_1 J_2 (J_1 + J_2) }}
		{i \sqrt{\pi} \left[ g^4 + 3 J_1^2 (J_1 + J_2)^2 - 2g^2 (J_1 + J_2) (2 J_1 + J_2) \right]  t^{1/2}} 
														e^{-i (J_1 + J_2) t} 
\label{A4.NZ}
\eeq
\end{widetext}
where we have made use of the integral $\int_0^\infty ds \; e^{-s} s^{-1/2} = \sqrt{\pi}$.
Combining this with the comparable result for the lowest band edge contribution $A_1 (t)$, we finally get
\beq
  \Abr (t) 
  	\approx - \frac{2 g^2 \alpha_\textrm{NZ,1} }{\sqrt{\pi} \ t^{1/2}}  
			\sin \left[ \left( J_1 + J_2 \right) t + \pi/4 \right]
	,
\label{1A.NZ.outer}
\eeq
which yields an oscillating $1/t$ evolution in the survival probability $P(t)$.  Here, the coefficient $\alpha_\textrm{NZ,1}$ is given by
\beq
  \alpha_\textrm{NZ,1}
  	= \frac{\sqrt{2 J_1 J_2}}
		{\sqrt{(J_1 + J_2)} X}  
\eeq
This is the equivalent of the near zone dynamics from Refs. \cite{GPSS13,GNOS19}.  Notice that Eqs. (\ref{1A.FZ.outer}) and (\ref{1A.NZ.outer}) reveal that the oscillations in the survival probability persist in Regions II and III, even though the outer bound states have vanished from the spectrum in these cases.


\section{Analysis of the $| \psi_\textrm{FM} (t) \ket$ contribution to the small $g$ dynamics on the lattice}\label{app:1A.FM}
\subsection{Dynamics on the $B$ sublattice}\label{sec:dyn.Bsublattice}
Here we work out the free motion-like contribution to the weak-coupling dynamics presented in Sec. \ref{sec:1A.dynamics.topological}.  Our simulations revealed this effect was the primary means of populating sites on the $B$ sublattice; hence, we will now study the amplitude
\beq
  A_{n,B} (t) = \bra n, B | e^{-i H t} | q \ket
\label{nB.amp}
\eeq
for $g \ll J_1^2 - J_2^2$.
As a first step, we will apply the modified resolution of unity
\beq
  \hat{I} = | \psi_a \ket \bra \psi_a | +  | \psi_s \ket \bra \psi_s |  
  	+  \sum_{\sigma=\pm} \int_{-\pi}^\pi \frac{dk}{2\pi} | \psi_{k,\sigma} \ket \bra \psi_{k,\sigma} |
	,
\label{unity.sa}
\eeq
written in terms of the symmetric and anti-symmetric combination states $ | \psi_{s,a} \ket$.

Inserting this into Eq. (\ref{nB.amp}) we have
\beqa
  A_{n,B} (t) & = & \bra n, B | e^{-i H t} | \psi_s \ket 		  \nonumber  \\
  & & 	+  \sum_{\sigma=\pm} \int_{-\pi}^\pi \frac{dk}{2\pi} 
			e^{-iE_{k,\sigma}t} \bra n, B | \psi_{k,\sigma} \ket \bra \psi_{k,\sigma} | q \ket
			\nonumber	\\
  & \equiv & A_{bs} (t) + A_c (t)
	,
\label{nB.amp.1}
\eeqa
since $\bra \psi_a | q \ket = 0$.  The first term $A_{bs} (t)$ is clearly related to the bound states while $A_c (t)$ appears as an integration over the continuum states.
Applying Eq. (\ref{psi.sa.time}) and the expansion for $ | \psi_s \ket $ from (\ref{psi.s.g})
allows us to approximate $A_{bs} (t)$ 
as
\beqa
  A_{bs} (t)
  	\approx \cos ( E_+ t )  \left( - \frac{g J_2}{J_1^2} \right)  \left( - \frac{J_2}{J_1} \right)^{n-1} 
  	.
\label{nB.bs}
\eeqa

To analyze the continuum component $A_c (t)$, we next write the amplitudes appearing under the integration.  The first of these is
\beqa
  \bra \psi_{k,\sigma} | q \ket
  &	= & \frac{g J_2}{E_k^2 -g^2}  \bra \psi_{k,\sigma} | 1, B \ket
					\\
  &	= & \frac{\sigma}{2 i} \frac{g J_2}{E_k^2 -g^2} \left( \sqrt{\frac{w_k {\tilde w}_{-k}}{w_{-k} {\tilde w}_{k}}}e^{ik}  
   		-  \sqrt{\frac{w_{-k} {\tilde w}_{k}}{w_{k} {\tilde w}_{-k}}} e^{-ik}\right),
		\nonumber
\eeqa
while the second amplitude is given by
\beq
  \bra n, B | \psi_{k,\sigma} \ket
  	= \frac{\sigma}{2 i} \left( \sqrt{\frac{w_k {\tilde w}_{-k}}{w_{-k} {\tilde w}_{k}}}e^{ikn}  
   		-  \sqrt{\frac{w_{-k} {\tilde w}_{k}}{w_{k} {\tilde w}_{-k}}} e^{-ikn}\right) 
	.
\eeq
Applying these two expressions and simplifying a bit, we can write the continuum component as
\beq
 A_c (t) = - 2 i g J_2 \int_{-\pi}^\pi \frac{dk}{2\pi} \ \frac{ \sin k \; \cos (E_{k,+} t)} {w_{-k} {\tilde w}_{k}} e^{ikn}
 	.
\label{Ac.3}
\eeq
Changing the integration variable to $\lambda = e^{ik}$, the integration path becomes the unit circle directed counterclockwise:
\beq
  A_c (t) 
  	= - \frac{g J_2 J_1}{2 \pi i \left( J_1^2 - g^2 \right)} 
		\oint d \lambda \frac{\left( \lambda - \lambda^{-1} \right) \cos \left(E_{k,+}(\lambda ) t \right) \lambda^n}
			{\left( J_2 \lambda + J_1 \right) \left( \lambda -\lambda_b  \right)
				}
\label{Ac.lambda}
\eeq
where
\beq
\lambda_b=-\frac{J_1 J_2}{J_1^2 - g^2}
\eeq
The integrand in Eq. (\ref{Ac.lambda}) contains the following singularities. There is a pole at $\lambda=-J_1/J_2$ associated with the zero mode, corresponding to $w_{-k}=0$; we will ignore this pole because it is outside the unit circle for $J_1>J_2$. There is also a  pole at $\lambda_b$ inside the unit circle, corresponding to ${\tilde w}_k=0$ and $E_{k,\pm} =E_{\pm}$ (the bound-state energies). Finally, there is an essential singularity at $\lambda=0$ associated with the factor $\cos (E_{k,+} t)$.  We will consider the bound-state and essential singularity contributions separately below.  This latter one yields the free motion-like evolution.


\subsection{Pole from the bound-state energies}\label{sec:dyn.bound}
\begin{figure}
 \includegraphics[width=0.45\textwidth]{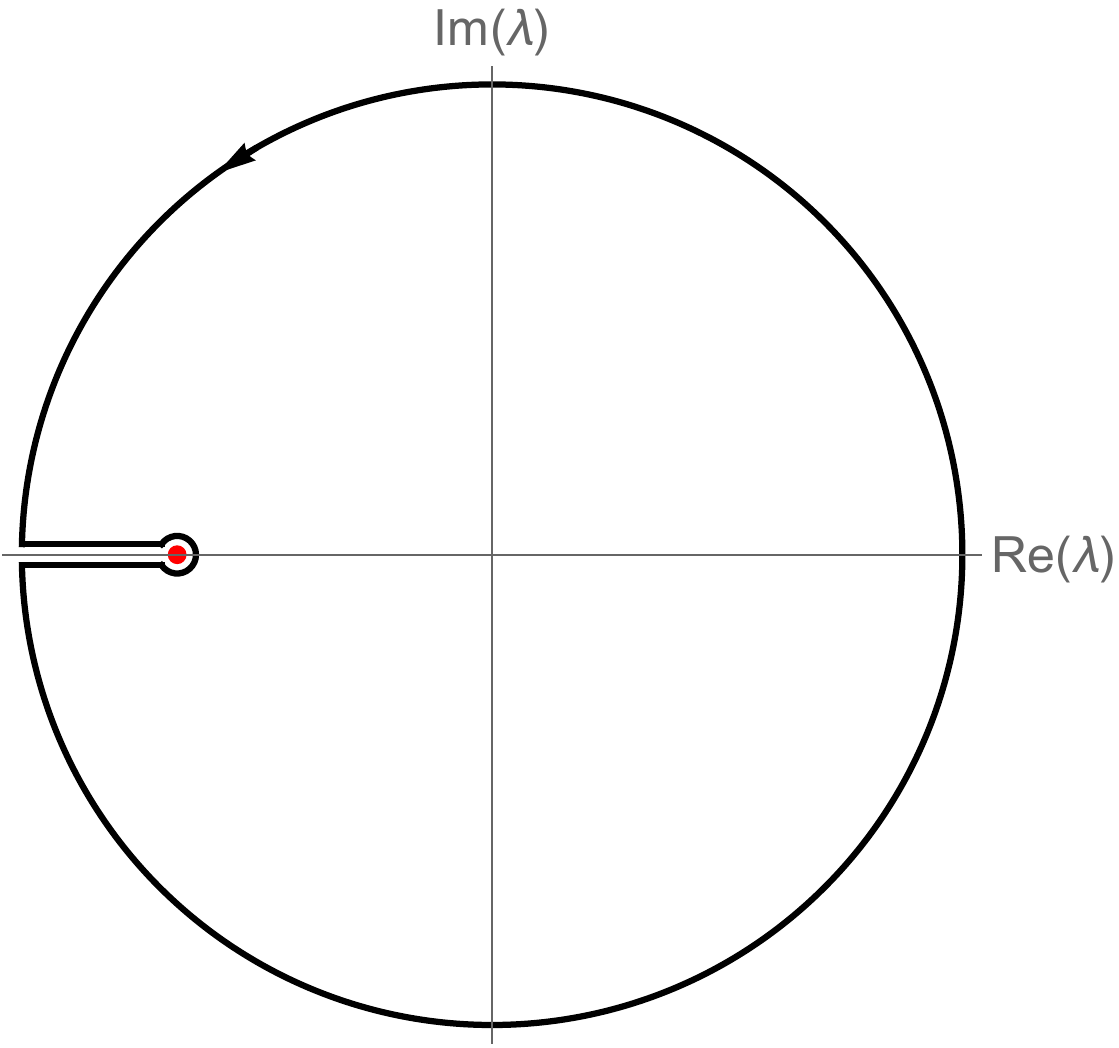}
\hfill
\\
\caption{ Integration contour in the complex $\lambda$ plane giving the amplitude $  A_{n,B} (t)$. The contour consists of the large (unit) circle, a small circle surrounding the pole at $\lambda=\lambda_b$, and two parallel lines joining both circles.  The parallel lines are infinitesimally close. The  pole at $\lambda=\lambda_b$ is marked by the red dot.
}
 \label{fig:contourfreem}
 \end{figure}

Noting that $A(t=0) = \bra n, B | q \ket = 0$ allows us to write
\beq
  A_{bs} (0) = - A_c (0),
\eeq
thus
\beq
    A_{bs} (t)=  - A_c (0) \cos (E_+ t) 
  . 
\eeq
The right-hand-side is precisely the residue of the integral in Eq. (\ref{Ac.lambda}) at $\lambda=\lambda_b$, which is obtained by an integration along a  small clockwise contour surrounding $\lambda_b$. Therefore the complete amplitude $  A_{n,B} (t)$ may be written as the integral in Eq. (\ref{Ac.lambda}) where the unit circle is replaced by the modified contour shown in Fig. \ref{fig:contourfreem}.

\subsection{Free motion contribution}\label{sec:dyn.cont}

The factor $\cos \left(E_{k,+}(\lambda ) t \right)$ from Eq. (\ref{Ac.lambda}) indicates the presence of an essential singularity in the integration at $\lambda=0$ because for $\lambda\to 0$, $E_{k,+}\to \sqrt{J_1J_2 \lambda^{-1}}$.  To evaluate this contribution we found it was easier to return to the form of the integral in Eq. (\ref{Ac.3}), keeping in mind that the other singularity at  ${\tilde w}_k=0$ is cancelled by the bound-state contribution as shown in the previous subsection. Hereafter we will ignore this other singularity. As a first step in the present analysis, we will apply a linearization of the dispersion $E_{k,+}$.


\subsubsection{Linearization of $E_k$}\label{sec:dyn.cont.group}

Let us split the $\cos \left(E_{k,+}(\lambda ) t \right)$ factor in Eq. (\ref{Ac.3}) 
into two exponential terms in order to write
\beq
  A_c (t) = A_{c,+} (t) + A_{c,-} (t)
\eeq
in which
\beq
  A_{c,\pm} (t)
  	=  - i g J_2 \int_{-\pi}^\pi \frac{dk}{2\pi} \ \Phi (k) e^{i \left( kn \pm E_k t \right)}
\label{Ac.5}
\eeq
and
\beq
  \Phi (k) \equiv  \frac{ \sin k} {w_{-k} {\tilde w}_{k}}
  	.
\eeq

In the usual wave packet analysis, one assumes that the integrand function $\Phi(k)$ is peaked at a particular $k_p$ and that the dispersion $E_k$ can be approximated through a linear expansion in $(k-k_p)$ in the vicinity of the peak.  The group velocity is then determined as the coefficient of the linear term in this expansion.  If we follow this method in the present case, we find an expression for the group velocity that gives a decent--- but not exact agreement with numerical simulations.


\begin{figure}
 \includegraphics[width=0.45\textwidth]{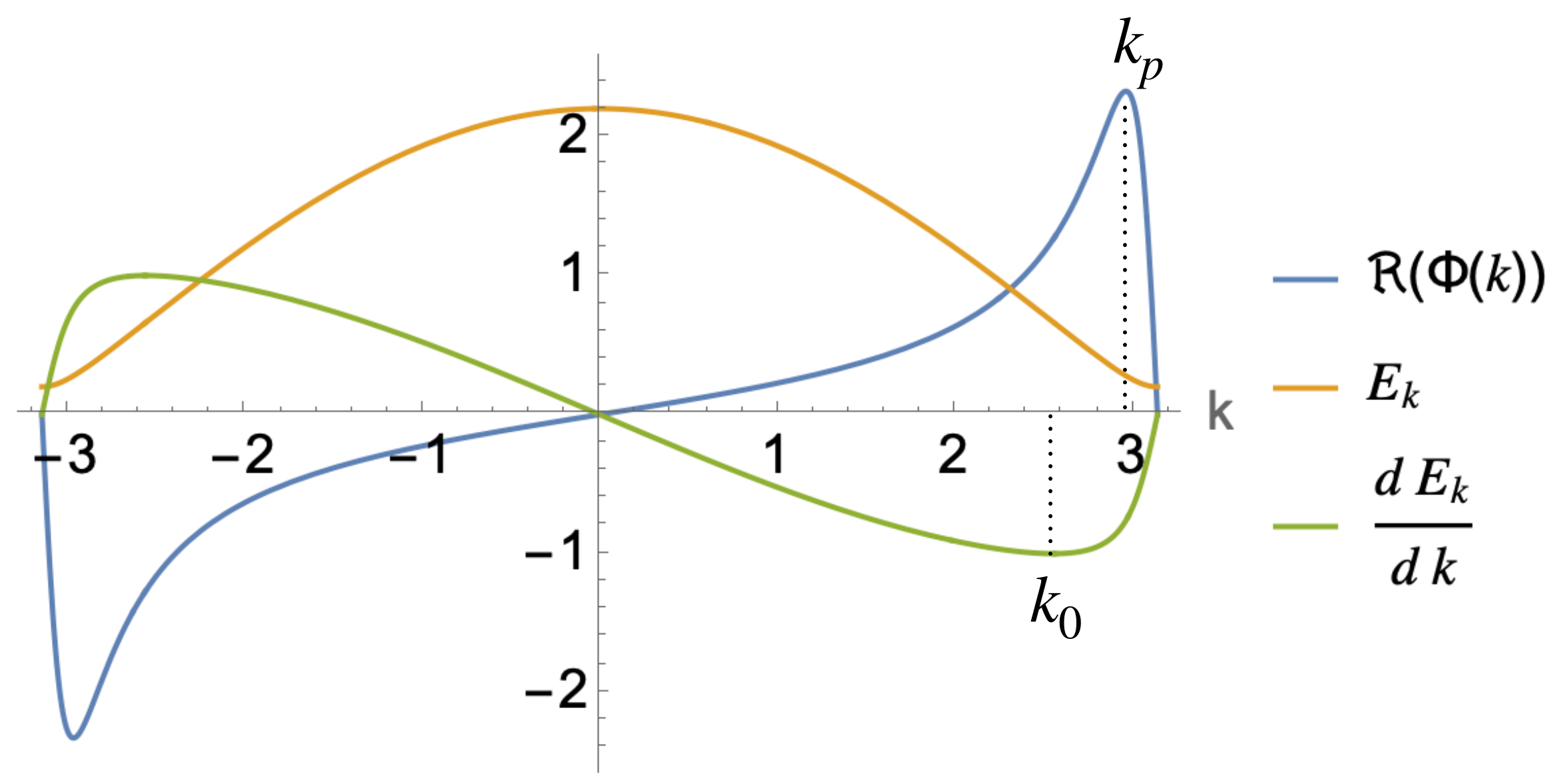}
\hfill
\\
\caption{ Graph of $E_k$, $dE_k/dk$ and the real part of $\Phi(k)$ vs. $k$ showing the location of the peak of $\Phi(k)$ at $k=k_p$ and the point $k_0$ where  $E_k$ is linearized (that is, where $d^2E_k/dk^2 = 0$).  Parameters are $J_1=1.2$, $J_2=1$ and $g=0.1$.
}
\label{fig:WavePacketAnalysis}
\end{figure}
However, we found that we can improve agreement with the simulations with a slight variation on the usual method.  We note in figure \ref{fig:WavePacketAnalysis} that $E_k$ is not linear at the peak of $\Phi(k)$, because the fist derivative changes rapidly around the peak.  Hence, we instead {\it impose} the linearization on $E_k$ by setting $\frac{\partial^2 E_{k}}{\partial k^2}=0$, which is equivalent to solving  a quadratic equation in $\cos k$. Solving this, we find the relevant solution as $k=k_0$, where
\beq
  \cos k_0 = - \frac{J_2}{J_1}
\eeq
(the second solution
  $\cos k_0 = - J_1/J_2$    
is spurious).

Around this point $E_k$ is approximately linear in $k$  because $\partial^2 E_k / \partial k^2 |_{k=k_0} = 0$. Therefore as a rough approximation we write 
\beq
  E_k \approx E_{k_0} + \left( k - k_0 \right) E_{k_0}' 
	.
\label{Ek.linear}
\eeq
where
\beq
E_{k_0}' \equiv \left(\frac{\partial E_k}{\partial k}\right)_{k=k_0}.
\eeq
From here, one might expect that $E_{k_0}' = - J_2$ should give the group velocity.  However, the propagation observed in Fig. \ref{fig:osc11} is clearly positive, not negative.  We will see the resolution of this issue shortly.



\subsubsection{Wave packet analysis}\label{sec:dyn.cont.wave}

For small $g$, the function $\Phi(k)$ is approximately antisymmetric: $\Phi(k) \approx -\Phi(-k)$. Therefore we write the continuum integral as 
\beq
  A_{c,\pm} (t)
  	=  - i g J_2 \int_{0}^\pi \frac{dk}{2\pi} \ \Phi (k)  e^{\pm i E_k t } \left(e^{i kn} - e^{-i kn}\right)
\label{Ac.7}
\eeq
Applying our linearization of $E_k$ from Eq. (\ref{Ek.linear}) allows us to write
\beqa
 && A_{c,\pm} (t)
  	= - i g J_2 e^{i\left(k_0n \pm  E_{k_0}t\right)}e^{-i k_0\left(n  \pm E_{k_0}'  t\right)} \nonumber\\
	&& \times \int_{0}^\pi \frac{dk}{2\pi} \ \Phi (k) 
  					e^{\pm i E_{k_0}' kt } \left(e^{i kn} - e^{-i kn}\right)
	,
\label{Ac.6}
\eeqa
which now appears in the standard wave packet form.  The factor 
$ e^{i \left( k_0 n \pm E_{k_0} t \right)}$ out front of the integral determines the phase velocity, while the factor including the integral determines the enveloping function (or `packet') of the wave, associated with the group velocity evolution.

We can easily obtain a lowest-order approximation for the integral in Eq. (\ref{Ac.6}) by expanding $\Phi (k)$ in powers of $k - k_0$, to write $\Phi (k) = \Phi_0 + \mathcal{O} \left( k - k_0 \right)$, in which
the lowest-order term is given by
\beq
  \Phi_0 = \frac{J_1}{\left( J_1^2 - g^2 \right) \sqrt{J_1^2 - J_2^2} - i g^2 J_2}
	.
\eeq
The integration in Eq. (\ref{Ac.6}) then becomes elementary, yielding
\beqa
  A_{c,\pm} (t)
  &	= & - \frac{g J_2 \Phi_0}{2 \pi } 
		e^{i \left( k_0 n \pm E_{k_0} t \right)} 	e^{-i k_0\left(n  \pm E_{k_0}'  t\right)}	
									\label{Ac.7}   	\\
  & &		\times \left[ \frac{e^{i \pi \left(n \pm E_{k_0}'t\right)} -1}{n \pm E_{k_0}' t} +  \frac{e^{-i \pi \left(n \mp E_{k_0}'t\right)} -1}{n \mp E_{k_0}' t}\right]		\nonumber
	.
\eeqa

To determine the conditions in which Eq. (\ref{Ac.7}) might give a non-negligible contribution to the evolution, consider that $| A_{c,\pm} (t)|^2$ appears to give $| A_{c,\pm} (t)| \sim g^2$, which we are neglecting.  Hence, to have a significant impact in the evolution, one should find conditions such that the factor $g$ in the numerator of Eq. (\ref{Ac.7}) cancels.  To achieve this, note that when we take the square modulus for the probability, we would need to include the overall factor $| \Phi_0 |^2$, which can be approximated as
\beqa
  \left| \Phi_0 \right|^2
  	\approx  \frac{1}{J_1^2 \left( J_1^2 - J_2^2 \right)}
		+ \mathcal{O} (g^2)
  	.
\eeqa
Next, let us quantify `smallness' of the band gap $J_1 - J_2$
by rewriting the parameter $J_2$ in terms of $J_1$ as
\beq
  J_2 \equiv J_1 + \delta
  	.
\eeq
In the case that the band gap becomes narrow, we would have $\delta \ll 1$ becomes a small quantity, and hence we could approximate
  $\left| \Phi_0 \right|^2
 \sim 1/2 J_1 \delta $
  	.
This would in turn yield
  $A_{c,\pm} (t)
  	\sim g / \delta$
	.
Hence, if $\delta \sim g$, we would expect the amplitude of $A_{c} (t)$ to become non-negligible, consistent with  the simulations presented in the main text.

Finally, we note that the denominators $\left( n + E_{k_0}' t \right)$ of Eq. (\ref{Ac.7}) give the dominant contribution to the evolution (recall that $E_{k_0}'= - J_2 $ is negative).  Further, the form of the exponential factors inside the brackets $\left[  \dots \right]$ in Eq. (\ref{Ac.7}) now makes clear that we should choose
\beq
  v_g = - E_{k_0}' = J_2
  	,
\eeq
yielding the anticipated {\it positive} expression for the group velocity $v_g$. The free motion contribution can be seen in Figs. \ref{fig:osc15}--\ref{fig:osc11} as the straight tilted line that starts from the origin.

\subsection{Dynamics on the $A$ sublattice}\label{sec:dyn.Asublattice}
Finally, we consider the amplitude
\beq
  A_{n,A} (t) = \bra n, A | e^{-i H t} | q \ket
\label{nA.amp}
\eeq
for $g \ll J_1^2 - J_2^2$.

Inserting a complete set of states we have
\beqa
  A_{n,A} (t) & = & \bra n, A | e^{-i H t} | \psi_a \ket 		  \nonumber  \\
  & & 	+  \sum_{\sigma=\pm} \int_{-\pi}^\pi \frac{dk}{2\pi} 
			e^{-iE_{k,\sigma}t} \bra n, A | \psi_{k,\sigma} \ket \bra \psi_{k,\sigma} | q \ket.
			\nonumber	\\
\label{nA.amp.1}
\eeqa
Similar to the amplitude for the $B$ sites, the second term (continuum contribution) is linear in $g$ in the lowest-order approximation. However, the first term (bound-state contribution) is of zeroth order in $g$. Hence in the lowest-order approximation we neglect the continuum contribution and use Eq. (\ref{psi.sa.time}) to obtain
\beqa
  A_{n,A} (t)  &\approx&  \bra n, A | e^{-i H t} | \psi_a \ket  \nonumber\\
  &=& -i \sin(E_+ t) \frac{\sqrt{J_1^2-J_2^2}}{J_1}\left(-\frac{J_2}{J_1}\right)^{n-1}.    
  \label{nA.amp.2}
\eeqa
The above expression corresponds to the horizontal dark-blue stripes seen in Figs. \ref{fig:osc15}--\ref{fig:osc11}. The continuum contribution gives a small ``free-motion'' component, which we can analyze as follows (for $n>1$):
\beqa
 &&  A_{c,A} (t)  =  \sum_{\sigma=\pm}\int_{-\pi}^\pi \frac{dk}{2\pi} 
			e^{-iE_{k,\sigma}t} \bra n, A | \psi_{k,\sigma} \ket \bra \psi_{k,\sigma} | q \ket
			\nonumber	\\
			&=&	 \sum_{\sigma=\pm} \int_{-\pi}^\pi \frac{dk}{2\pi} 
			e^{-iE_{k,\sigma}t} \frac{1}{2i}\left(\sqrt{\frac{{\tilde w}_{-k}}{{\tilde w_{k}}}}e^{ikn} - \sqrt{\frac{{\tilde w}_{k}}{{\tilde w_{-k}}}} e^{-ikn} \right)\nonumber\\
			&\times&   \frac{\sigma}{2 i} \frac{g J_2}{E_k^2 -g^2} \left( \sqrt{\frac{w_k {\tilde w}_{-k}}{w_{-k} {\tilde w}_{k}}}e^{ik}  
   		-  \sqrt{\frac{w_{-k} {\tilde w}_{k}}{w_{k} {\tilde w}_{-k}}} e^{-ik}\right)\nonumber\\
		&=& -2 g J_2 \int_{-\pi}^\pi \frac{dk}{2\pi} 
			\frac{\sin\left(E_{k,+}t\right)\sin(k)}{{\tilde w}_k E_{k,+}} e^{ikn} \label{nA.amp.3}
\eeqa
The integral in the last line has a pole at ${\tilde w}_k=0$, corresponding to the bound-state energy pole. The residue at this pole gives a small correction to the pole contribution in Eq. (\ref{nA.amp.2}). Other than this, the integral has a form similar to the integral in Eq. (\ref{Ac.3}) so it gives a freely-propagating wave packet with group velocity $v_g=J_2$ for the $A$ sites, similar to the freely-propagating wave packet for the $B$ sites in the previous subsection.

\end{document}